\documentclass[reqno,12pt]{article}
\usepackage{amsmath,amsfonts,amssymb,amsthm,amstext,amscd,eucal,xcolor}
\usepackage[all]{xy}
\usepackage{epsfig}
\usepackage{color}
\usepackage{graphicx}
\usepackage[active]{srcltx}
\makeatletter \@addtoreset{equation}{section}

\makeatletter\renewcommand\section{\@startsection {section}{1}{\z@}%
                                   {-3.5ex \@plus -1ex \@minus -.2ex}%nn
                                   {2.3ex \@plus.2ex}%
                                   {\normalfont\large\bfseries}}
\renewcommand\subsection{\@startsection{subsection}{2}{\z@}%
                                     {-3.25ex\@plus -1ex \@minus -.2ex}%
                                     {1.5ex \@plus .2ex}%
                                     {\normalfont\bfseries}}

\newcommand{\be}{\begin{equation}}
\newcommand{\ee}{\end{equation}}
\newcommand{\bea}{\begin{eqnarray}}
\newcommand{\eea}{\end{eqnarray}}
\newcommand{\bse}{\begin{subequations}}
\newcommand{\ese}{\end{subequations}}
\newcommand{\bi}{\begin{itemize}}
\newcommand{\ei}{\end{itemize}}

\newcommand{\beq}{\begin{eqnarray}}
\newcommand{\eeq}{\end{eqnarray}}

\newcommand{\nn}{\nonumber}

\def\s2s1{S$^2\times$S$^1$ }
\def\Label#1{\label{#1}%
  \smash{\hbox to0pt{\raise1ex\hbox{\tiny[#1]}\hss}}}
\def\noLabels{\let\Label=\label}
\def\nobbibitem{\let\bbibitem=\bibitem}

\begin{document}
\baselineskip 18pt%
\begin{titlepage}
\vspace*{1mm}%
\hfill%
%\vspace*{15mm}%
\hfill
\vbox{
    \halign{#\hfil         \cr
arXiv:yymm.nnnn\cr
         IPM/P-2014/nnn  \cr
          } % end of \halign
      }  % end of \vbox
%\vspace*{20mm}
\begin{center}
{\Large{\textbf{More on Five Dimensional EVH Black Rings}}}
\vspace*{8mm}

{ Ahmad Ghodsi\footnote{a-ghodsi@ferdowsi.um.ac.ir}$^{,a}$, Hanif Golchin\footnote{hanif.golchin@stu-mail.um.ac.ir}$^{,a}$ and M.M. Sheikh-Jabbari\footnote{jabbari@theory.ipm.ac.ir}$^{,b}$}\\
\vspace*{0.4cm}
{$^a$ \it Department of Physics, Ferdowsi University of Mashhad, \\
P.O. Box 1436, Mashhad, Iran}\\
{$^b$ \it School of Physics, Institute for Research in Fundamental Sciences (IPM),\\
 P.O.Box 19395-5531, Tehran, Iran}
\vspace*{1.0cm}
%and\\
%{ School of Physics,
% Institute for research in fundamental sciences (IPM), \\
% P.O. Box 19395-5531, Tehran, Iran.
%}\\\vspace*{1.5cm}
\end{center}

\begin{abstract}
In this paper we continue our analysis of arXiv:1308.1478[hep-th] and study  in detail the parameter space of three families of doubly spinning black ring solutions:  balanced black ring,  unbalanced ring  and  dipole-charged balanced black rings. In all these three families the Extremal Vanishing Horizon (EVH) ring appears in the vanishing limit of the dimensionful parameter of the solution which measures the ring size. We study the near horizon limit of the EVH black rings and for all three cases we find a (pinching orbifold) AdS$_3$ throat with the AdS$_3$ radius $\ell^2=8 G_5 M/(3\pi)$ where $M$ is the ring mass and $G_5$ is the 5d Newton constant. We also discuss the near horizon limit of near-EVH black rings and show that the AdS$_3$ factor is replaced with a generic BTZ black hole. We use these results to extend the EVH/CFT correspondence for black rings, a 2d CFT dual to near-EVH black rings.
\end{abstract}

%the known five dimensional black ring solutions which admit the Extremal Vanishing Horizon (EVH) limit. We discuss

\end{titlepage}
\addtocontents{toc}{\protect\setcounter{tocdepth}{2}}
\tableofcontents

\section{Introduction}
Five dimensional asymptotic flat vacuum Einstein gravity solutions are in the form of Myers-Perry black holes \cite{MP} or black rings \cite{Emparan:2001wn,Pomeransky:2006bd, Emparan:2008eg}.  These solutions form a three parameter family, specified by mass and two angular momenta. In different regions of the parameter space of these solutions  we can have black hole solutions with $S^3$ horizon topology \cite{MP}, or black ring solutions with $S^2\times S^1$ horizon topology \cite{Emparan:2001wn,Pomeransky:2006bd, Emparan:2008eg}. There is a region in this parameter space where black holes and rings can both exit \cite{Elvang-Emparan-2003}. In this overlapping region where the hole and the ring have the same mass and angular momenta, the black hole solution has a larger entropy than the ring  and so it is  expected to be a more stable configuration.

Black rings can be balanced or unbalanced. In the balanced case constructed and discussed in \cite{Emparan:2001wn,Pomeransky:2006bd}, expanding around north and south poles of topologically $S^2$ part of  the horizon we get a 2d flat space without any deficit angle or conical singularity. In a different viewpoint, in the balanced case the centrifugal force from the angular momentum along the ring is tuned to precisely balance off the tension and self-gravitation of the ring \cite{Elvang-Emparan-2003, Emparan-Obers-2007, ur}.
However, in the unbalanced case \cite{Elvang-Emparan-2003} (\emph{cf.} discussions in section \ref{sec-3}), if we adjust the expansion around the north pole of the topologically $S^2$ part of the horizon to be a flat $R^2$, the expansion around the south pole will show a deficit (or excess) angle and we have a conical disk \cite{A-R-T-papers}. For the unbalanced rings the mass gets an additional contribution from the pressure (tension) of the conical (defect) disk \cite{Elvang-Emparan-2003, ur, A-R-T-papers}. Due to the contribution of this tension the first law of thermodynamics and the Gibbs free energy for unbalanced rings has an extra term which vanishes in the balanced case  \cite{{A-R-T-papers},Chen:2011jb}.
The parameter space of unbalanced rings is hence four dimensional, mass, two spins and the unbalance parameter. The balanced rings solutions correspond to a three dimensional subspace on this parameter space.

The 5d Einstein-Maxwell-dilaton theory also admits charged black ring solutions. There are in particular black rings with conserved dipole charges, while having vanishing electric or magnetic charge \cite{dipole, Yazadjiev, Chen:2012kd}. The dipole charges can be defined as Noether-Wald conserved charge \cite{Sudarsky-Wald} and appear in the first law of black hole thermodynamics for the black rings \cite{Horowitz-Copsey}. The (balanced) dipole-charged black rings form a four parameter family of solutions, describing the mass, two spins and the dipole charge.

In the family of balanced, unbalanced and (neutral but) dipole charged black rings we have geometries with vanishing surface gravity  and degenerate horizon, the extremal rings. Within the family of extremal rings, there exist a ``singular'' region where the horizon area (and hence the Bekenstein-Hawking entropy) vanishes. These Extremal Vanishing Horizon (EVH) rings are what we will focus on in this work. One of the motivations to study the EVH rings is that, as discussed in our previous work on the topic \cite{Ghodsi:2013soa}, the EVH points are generically in the part of the parameter space of the ring solutions where the hole and ring solutions can coexist \cite{Elvang-Emparan-2003} and the hole-ring transition can be studied.

Despite of being singular, the EVH black hole/rings, have their own interesting features which makes their analysis worthwhile. Many different examples of EVH black holes in various dimensions and various asymptotics and theories have been studies, see \cite{SheikhJabbaria:2011gc,EVH-examples,deBoer:2011zt} for an incomplete list.  To be more precise, EVH black holes/rings are defined as black objects with $A_H, T_H\to 0$ limit while $T_H/A_H$ is kept fixed, where $A_H$ is the horizon area and $T_H$ is the Hawking temperature \cite{SheikhJabbaria:2011gc}. Moreover, the vanishing of horizon area should come from vanishing of a one-cycle on the horizon \cite{Johnstone:2013ioa}. In all these various examples it has been observed that the near horizon limit of EVH black holes leads to an AdS$_3$ throat. This AdS$_3$ factor is, however, a pinching orbifold of AdS$_3$ \cite{deBoer:2010ac}. Moreover, one may consider ``excitations'' of these EVH black holes to near-EVH black holes. As one may expect,  in the near horizon limit of the near-EVH black holes the  (pinching) AdS$_3$ is then excited to (pinching) BTZ. This nice feature prompts the idea that one may be able to study the low energy excitations of EVH black holes focusing on their near horizon geometry and their excitations. This was actually what was proposed as EVH/CFT correspondence \cite{SheikhJabbaria:2011gc}, that low energy excitations around an EVH black hole is described by a 2d CFT dual to the AdS$_3$ factor appearing in its near horizon geometry.\footnote{Although the near horizon geometry, like the original EVH solution, has curvature singularity, this singularity is of the ``good type'' in the terminology of \cite{Gubser}. That is, if we reduce the gravity theory on the ansatz given by the near horizon EVH geometry (to obtain an AdS$_3$ gravity), the reduced theory does not involve any singularity. In other words, the low energy excitations appearing in near-EVH black holes do not probe the singularity, they are completely captured in the excitations of the AdS$_3$ throat.}

There are also EVH black rings, falling precisely into the definition of EVH black holes given above \cite{Ghodsi:2013soa}. As is expected and we will show explicitly in this work, for EVH rings the vanishing of horizon area should come from vanishing of a circle on the topologically $S^2$ part of the geometry and the ring size remains finite. We will show that, like EVH black holes, we get a (pinching) AdS$_3$ in the near horizon limit of the EVH black rings.
As we will see this AdS$_3$ throat consists of the radial $r$ and time direction $t$ of the original ring and the vanishing circle on the topologically $S^2$ part of the geometry, the ring circle is transverse to the AdS$_3$ throat. Moreover, in the near-EVH black rings this AdS$_3$ factor is excited to a BTZ black hole.

In this work we will give a full account of balanced and unbalanced doubly spinning rings and doubly spinning dipole charge rings and where in their parameter space they become EVH. As we will show for both balanced and unbalanced cases the near-EVH excitations appear as generic BTZ black hole excitation on the AdS$_3$ throat. We discuss how the unbalancing factor and the dipole charge appear in the near horizon geometry and its excitation. We discuss the EVH/CFT correspondence as the dual 2d CFT describing low energy excitations of the EVH rings. Among other things, we also discuss how the EVH Myers-Perry black hole and an EVH ring of similar mass and spin can be distinguished from this dual CFT viewpoint.

The organization of the paper is as follows. In section \ref{sec-2}, we revisit and extend the EVH Pomeransky-Sen'kov black ring solution
analyzed in \cite{Ghodsi:2013soa} and study the most general region in the parameter space which the EVH ring solutions exist.
In section \ref{sec-3}, we investigate the parameter space of the unbalanced double rotating black ring \cite{Elvang-Emparan-2003,A-R-T-papers,Chen:2011jb}
and specify the region corresponding to EVH unbalanced rings. We also  discuss the near horizon geometry of these EVH rings. The parameter space of the double
rotating dipole black ring  solution \cite{Chen:2012kd}, its EVH regions and the corresponding near horizon geometry is studied in section \ref{sec-4}.
We find that in the EVH limit the dipole charge is irrelevant. We discuss the EVH/CFT correspondence for the mentioned EVH rings in section \ref{sec-5}.
Last section is devoted to concluding remarks.
%%%%%%%%%%%%%%%%%%%%%%%%%%%%%%%%%%%%%%%%%%%%%%%%%%%%%%%%%%%%%%%%%%%%%%%%%%%%%%%%%
\section{Balanced EVH   Pomeransky-Sen'kov black ring}\label{sec-2}
Neutral double rotating black ring \cite{Pomeransky:2006bd} (DRBR) is an asymptotically flat vacuum solution of  Einstein gravity with the horizon topology  $S^1\times S^2$. In \cite{Ghodsi:2013soa} we studied the parameter space of this solution, focusing on the  Extremal Vanishing Horizon (EVH) limit. This is a limit where the Hawking temperature $T_H$ and the horizon area $A_H$ of the solution vanishes:
\be \label{evhc}
A_H \to 0\,, \qquad T_H \to 0\,, \qquad \frac{T_H}{A_H}\to finite\,.
\ee
We showed that condition (\ref{evhc}) can be satisfied in the parameter space of DRBR and discussed  a specific EVH point in this parameter space. In this section we study the most general EVH region in the parameter space of DRBR, extending and generalizing discussions in \cite{Ghodsi:2013soa}.

Let us start with reviewing the solution. The DRBR line element is given by
\bea \label{DRBRmetr}
ds^2&=& -\frac{H(y,x)}{H(x, y)}\big(d t+\Omega(x, y)\big)^2-\frac{F(x, y)}{H(y,x)}d\psi^2-2\frac{J(x, y)}{H(y,x)} d \phi\, d\psi
\nn \\ &+&\frac{F( y,x)}{H( y,x)}d \phi^2+\frac{2 k^2 H(x, y)}{(x- y)^2(1-\nu)^2}\big(\frac{dx^2}{G(x)}-\frac{d y^2}{G(y)}\big)\,,
\eea
where $-1\leq x\leq1$ and $-\infty <y<-1$ and $\phi, \psi\in[0, 2\pi]$. The functions $F, G, H, J$ and $\Omega$ are defined as follows
\be\label{FGHJOm}
\begin{split}
F(x, y) &= \frac{2 k^2}{(x -  y)^2 (1 - \nu)^2} \Big( G(x) (1 -  y^2)\big(((1 - \nu)^2 - \lambda ^2)
(1 + \nu ) +  y \lambda (1 \!- \!\lambda ^2 + 2 \nu  - 3 \nu ^2)\big) \\
&+ G( y) (2 \lambda ^2  +  x \lambda ((1 - \nu )^2 + \lambda ^2)+ x^2\big((1 - \nu )^2 - \lambda ^2\big) (1 + \nu) \\
&+ x^3\lambda(1 - \lambda^2 - 3\nu^2 + 2\nu^3) -  x^4 (1 - \nu ) \nu (\lambda ^2 + \nu ^2 - 1))\Big) \,,\\
G(x)&=(1-x^2)(1+\lambda x+\nu x^2)\,, \\
 H(x, y) &= 1+\lambda ^2-\nu^2+2\lambda\nu (1-x^2) y+2x\lambda(1- y^2\nu^2)+ x^2  y^2 \nu(1-\lambda^2-\nu^2)\,,\\
J(x, y)&=\frac{2 k^2 (1-x^2) (1- y^2) \lambda  {\nu^\frac12}}{(x- y) (1-\nu)^2} \big(1+\lambda ^2 -\nu ^2 + 2 (x+ y) \lambda  \nu-x  y \nu (1-\lambda ^2-\nu^2)\big)\,,\\
\Omega(x, y)&=-\frac{2 k \lambda \big((1+\nu )^2-\lambda ^2\big)^\frac12}{H( y,x)} \big(\frac{1+ y}{1-\lambda +\nu}(1+\lambda -\nu +\nu (1-\lambda-\nu) y x^2+2\nu x(1- y))d\psi \\
&+\nu^\frac12  y(1-x^2)d\phi \big)\,.
\end{split}
\ee
In above solution  $k$  has dimension of length and is related to the radius of the ring circle which is parameterized by $\psi$.\footnote{We use the notation that coordinates $\phi$ and $\psi$ are interchanged compared to the original paper \cite{Pomeransky:2006bd}. This is the notation used in papers by Emparan et al.  The metric also has written  with the mostly plus signature.}
 On the other hand $\nu$ and $\lambda$ are two dimensionless parameters related to the rotations of the black ring around the $\phi$ and $\psi$ directions with
\be \label{ranges1}
 0\leq\nu<1\,, \qquad  2\sqrt{\nu}\leq\lambda<1+\nu\,,  \qquad  k>0\,.
\ee
$\nu$ controls the rotation around the $\phi$ direction, which parameterizes the circle on the topologically $S^2$ part of the horizon; in the $\nu=0$ we  recover the single rotating black ring of Emparan and Reall \cite{Emparan:2001wn}. The parabola $\nu=\lambda^2/4$ is where the black ring becomes extremal. The parameter space for solution (\ref{DRBRmetr}) is depicted in Fig(\ref{fig1}.b).

In \cite{Ghodsi:2013soa} we discussed that around the cusp  at $\nu=1, \lambda=2$\, (the ``collapsing region'' \cite{Elvang:2007hs})  Hawking temperature and  Beckenstein-Hawking entropy of this solution which are given by
\be \label{entem}
T_{H}=\frac{(y_h^{-1} - y_h) (1-\nu) \sqrt{\lambda^2 - 4 \nu}}{8\pi\, k\, \lambda (1+\nu +\lambda)} \,, \qquad\quad
S_{BH}=\frac{8 \pi^2 k^3 \, \lambda (1+\nu+\lambda)}{G_5(1-\nu)^2(y_h^{-1}-y_h)} \,,
\ee
satisfy the EVH condition provided that we scale $k$ appropriately. In the above
\be\label{yh-DRBR}
y_{h}=\frac{-\lambda + \sqrt{\lambda^2-4\nu}}{2\nu}\,,
\ee
is the outer horizon (larger root of $G(y)$) of the DRBR.
 One may then readily check that  the other physical quantities,  the mass, angular momenta and angular velocities
\footnote{For general values of the parameters we have the Smarr relation $M=\frac32\,(T_H S_{BH} + J_\phi \Omega_\phi + \Omega_\psi J_\psi)$\,.}
\bea \label{MJOm}
M&=&\frac {3\,\pi \,{k}^{2}\lambda}{G_5(1+\nu-\lambda)}\,, \qquad J_{\psi}=\frac {2\,\pi{k}^{3}\lambda \sqrt{ \left( 1+\nu \right) ^{2}-{\lambda}^{2}} \left( {\nu}^{2}+ \left( \lambda-6 \right) \nu+\lambda+1 \right) }{G_5 \left( 1-\nu \right) ^{2} \left( 1+\nu-\lambda \right) ^{2}}\,,\nn\\
\qquad J_{\phi}&=&\frac {4\,\pi \,\lambda \sqrt{\nu}\,{k}^{3} \sqrt{ \left(1+\nu \right) ^{2}-{\lambda}^{2}}}{G_5 \left( 1-\nu \right) ^{2} \left(1+\nu-\lambda \right) }\,, \qquad \qquad\qquad \quad \Omega_{\psi}=\frac{1}{2 k} \sqrt{\frac{1+\nu-\lambda}{1+\nu+\lambda}} \,,\nn\\
&& \qquad\quad \Omega_{\phi}=\frac{\lambda (1+\nu)-(1-\nu)\sqrt{\lambda^2 - 4\nu}}
 {4 k\, \lambda \sqrt{\nu}} \sqrt{\frac{1+\nu-\lambda}{1+\nu+\lambda}}\,,
\eea
remain finite at the $\nu=1,\ \lambda=2$ EVH point with the prescribed $k$ scaling \cite{Ghodsi:2013soa}. In this EVH point, $J_\phi$ vanishes while all the other thermodynamic quantities remain finite, and $3M=2\Omega_\psi J_\psi$.

In the collapsing point $\lambda=2,\ \nu=1$\, there is the coordinate transformations which transforms  DRBR to a generic (non-extremal) Myers-Perry (MP) black hole \cite{Elvang:2007hs}
\be \label{trans1}
x=-1+\frac{16 \sqrt{a} k^3 \cos^2\theta}{(a+b)^{3/2}(r^2-a\,b)}\,, \quad \qquad y=-1-\frac{16 \sqrt{a} k^3 \sin^2\theta}{(a+b)^{3/2}( r^2-a\,b)}\,,
\ee
where $a$ and $b$ are the rotation parameters of the MP black hole and are given by $a=\sqrt{2\tilde M} \sigma$ and $b=\sqrt{2\tilde M}(1-\sigma)$. In addition $\sigma$ and $\tilde M$ are defined by
\be \label{sM}
\sigma=\frac{1+\nu-\lambda}{(1-\nu)^2}\,,\qquad \qquad \tilde M=\frac{8k^2}{1+\nu-\lambda}\,.
\ee
We note that (\ref{trans1}) demonstrates two features of an EVH near horizon solution: For the cusp at $\lambda=2,\,\nu=1$, horizon is located at $y_h\sim-1$\, and recalling that we need to take $k\to 0$ to keep the mass finite, (\ref{trans1}) is essentially a near horizon expansion. Second, we should obtain a vanishing horizon solution due to {\it one} shrinking cycle on the horizon. To see this let us study  behavior of the metric on the horizon in the $\lambda=2,\,\nu=1\,$ limit. The EVH condition \eqref{evhc} is satisfied if one scales the parameters as
$$\nu=1-\hat \nu \epsilon\,,\qquad \lambda=1+\nu-\hat \lambda \epsilon^4,\qquad k=\hat k \epsilon^2\,.$$ Without inserting $x\sim y\sim -1$ as in (\ref{trans1}), the metric components on the horizon (constant $t$ and $y=y_h$), are
\be \label{HMcomp}
g_{xx}=\frac{-8\,\hat k^2}{\hat \nu(x-1)(x+1)^3}\epsilon^3\,,\qquad g_{\phi\phi}=\frac{8(1-x)\hat k^2}{(x+1)\hat \nu}\,\epsilon^3\,, \qquad g_{\psi\psi}=\frac{16\,\hat k^2}{\hat \lambda}\,.
\ee
It is obvious that there are {\it two} vanishing components in the  horizon metric unless $x\sim -1+\epsilon$. In other words, (\ref{trans1}) insures two necessary properties of an EVH near horizon solution.

\subsection{The EVH and near-EVH near horizon limit}

As pointed out in \cite{Ghodsi:2013soa}, one may formally check that for a generic point on the $\lambda=1+\nu$ line (and not just the $\nu=1$ on it), both $T_H,A_H\to 0$ and $T_H/A_H=finite$ conditions still hold. However, as depicted in Fig(\ref{fig1}.b) this line (shown as a dashed line) is not in the DRBR parameter space and the solution becomes singular on this line. Despite of this fact, given the ``EVH-type'' behavior of temperature and entropy, one can show that when we approach this dashed line from the left we indeed get an EVH ring for generic values of $\nu$. This is what we will establish here. To this end,
we consider $\lambda=1+\nu-\hat \lambda \epsilon^a$ expansion with $\hat\lambda>0$. The parameter $a$ controls how fast we are approaching the $\lambda=1+\nu$ line. As we show momentarily $a\geq 4$ corresponds to sitting at the EVH point while with $a=2$ we find a near-EVH ring.

In order to find the near horizon of the EVH DRBR, we define the parameters of the solution as follows
\be \label{parsca}
\lambda=1+\nu-\hat \lambda \epsilon^4\,, \qquad \quad k=\hat k \epsilon^2\,,
\ee
together with transformations (\ref{trans1}), accompanied by inserting the following scalings in the coordinates
\be \label{csca1}
r =\hat r \epsilon\,, \qquad  t=\frac{\hat t}{\epsilon}\,, \qquad \psi=\hat \psi+\Omega_{\psi}\,t\,,\qquad \phi=\frac{\hat \phi}{\epsilon}\,.
\ee
Using (\ref{parsca}) the mass and spins of the solution behave as $M\sim M_0+\delta M \epsilon^4\,,\, J_{\phi}\sim \delta J_{\phi}\epsilon^4$ and $J_{\psi}\sim J_{\psi_0} +\delta J_{\psi}\epsilon^4\,$. The near horizon metric can be obtained by taking the limit $\epsilon\to 0$
\bea \label{nhDRevh}
ds^2=\cos^2\theta \bigg[-\frac{\hat r^2}{\ell^2}\,d\hat t^2+\frac{\ell^2}{\hat r^2}\,d\hat r^2+\hat r^2 d\hat \phi^2\bigg]+\ell^2\left(\cos^2\theta\, d\theta^2+\tan^2\theta d\hat \psi^2\right),
\eea
where $\ell^2=\frac{8\hat k^2(1+\nu)}{\hat \lambda}$ is the AdS$_3$ radius which is related to the physical mass of the solution as $M_0=\frac{3\pi}{8}\ell^2$\,. In fact (\ref{nhDRevh}) is in the form of pinched AdS$_3$ due to the infinitesimal period of $\hat \phi\in [0, 2\pi\epsilon]$. We also note that \eqref{nhDRevh} is exactly the same geometry one find in the near horizon of EVH 5d Myers-Perry black hole \cite{Ghodsi:2013soa}. This is of course not a coincidence, as there are uniqueness theorems for 5d Einstein vacuum solutions with local $SO(2,2)$ invariance \cite{SO22-uniqueness}.

In order to find the near-EVH near horizon geometry, we should adjust how fast we approach the $\lambda=1+\nu$ line. We choose
\be \label{parscane}
\lambda=1+\nu-\hat \lambda \epsilon^2\,, \qquad \quad k=\hat k \epsilon\,,
\ee
where $\epsilon$ is the parameter defined in the near horizon scaling of the radial coordinate $r$, $r=\hat r \epsilon$.
We see that  mass and $J_{\phi}$ appear in the form
$M= M_0+\delta M \epsilon^2\,, J_{\phi}\sim \delta J_{\phi}\epsilon^2$ and  $J_{\psi}\sim J_{\psi_0} +\delta J_{\psi}\epsilon^2$, and
\be \label{shifts}
J_{\phi}\sim \epsilon^2\,, \qquad M-\frac{M_0}{J_{\psi_0}}J_{\psi}\sim \epsilon^2\,.
\ee
These charge scalings are compatible with the general expectation for EVH black holes \cite{Johnstone:2013ioa}.
The value of leading terms for $M, J_{\phi}$ and $J_{\psi}$ in the near-EVH limit are given by $(G_5=1)$
\be M_0=\frac{3\pi\hat k^2(1+\nu)}{\hat \lambda}\,,\quad \delta J_{\phi}=\frac{4\pi\hat k^3\sqrt{2\nu}(1+\nu)^{3/2}}{\sqrt{\hat \lambda} (\nu-1)^2}\,,\quad J_{\psi_0}=\frac{4\pi\hat k^3\sqrt{2}(1+\nu)^{3/2}}{\hat \lambda^{3/2}}\,.
\ee
After discussing the scaling of parameters we now consider the near horizon scaling of coordinates.  It is sufficient to use (\ref{parscane}) and (\ref{trans1}) together with the following rescalings
\be \label{csca2}
r =\sqrt{2\hat r-2\hat k^2\cos 2\theta}\,\, \epsilon\,, \qquad  t=\frac{\hat t}{\epsilon}\,, \qquad \psi=\hat \psi+\Omega_{\psi}\,t\,,\qquad \phi=\frac{\hat \phi}{\epsilon},\qquad \epsilon \to 0\,.
\ee

The near horizon metric after a redefinition $\hat r=\frac{\rho^2(1-\nu)+2\hat k^2(\nu+7)}{2(1-\nu)}$ can be written as
\be\label{nhDRnevh2}\begin{split}
ds^2=&\cos^2\!\theta \Big[-{f(\rho)} d\hat t^2+\frac{d\rho^2}{f(\rho)}+\rho^2 \big(d\hat\phi-\frac{8\hat k^2(1+\nu)\sqrt{\nu}}{\rho^2\ell(\nu-1)^2}d\hat t\big)^2\Big]+\ell^2\big(\cos^2\theta d\theta^2+\tan^2\theta d\hat \psi^2\big),\\
f(\rho)=&\frac{1}{\ell^2\rho^2}\big(\rho^2-\frac{4\hat k^2(\nu+1)^2}{(1-\nu)^2}\big)\big(\rho^2-\frac{16\hat k^2\nu}{(1-\nu)^2}\big)\,,\qquad
\ell^2= \frac{8\hat k^2(1+\nu)}{\hat \lambda}\,.
\end{split}
\ee
As we expected the AdS$_3$ throat of the geometry \eqref{nhDRevh} is now  replaced by  a generic (pinching) BTZ part which is a common feature for the near-EVH near horizon solutions. One may show that the mass and angular momentum of the BTZ is exactly equal to the near-EVH angular momentum and charges given in \eqref{shifts}. To see this we need to reduce the 5d gravity over the $\theta\hat\psi$ part of \eqref{nhDRevh} to obtain the Newton constant of the 3d AdS$_3$ gravity: $G_3=\frac{1}{\pi \ell^2}$ (see (\ref{grel})). Then, we use the standard formulas for the BTZ mass and spin
\be \label{mjbtz}
M_{BTZ}=\frac{\rho_+^2+\rho_-^2}{8\ell^2 G_3}=\frac{\pi \hat k^2(\nu^2+6\nu+1)}{2(\nu-1)^2}\,,\,\qquad J_{BTZ}=\frac{\rho_+ \rho_-}{4\ell G_3}=\frac{4\pi \hat k^3\sqrt{2\nu}(1+\nu)^{3/2} }{\sqrt{\hat \lambda}(\nu-1)^2}=\delta J_{\phi}\,.
\ee
There seems to be an $\epsilon^2$ factor difference between $M_{BTZ},\ J_{BTZ}$ and the expression in \eqref{shifts} which may be understood as follows \cite{Johnstone:2013ioa}: One power of $\epsilon$ comes from the fact that we have scaled $t$ and $\phi$ by $\epsilon$, \emph{cf}. \eqref{csca2}. The other factor comes from the pinching. (Recall that BTZ mass and spin are obtained as integrals over the $\hat\phi$ direction which is ranging over $[0,2\pi\epsilon]$.)

%%%%%%%%%%%%%%%%%%%%%%%%%%%%%%%%%%%%%%%%%%
\paragraph{The $\lambda=2, \nu=1$  limit.} It is quite natural to search for EVH black holes/rings among  extremal ones. This was what we did in \cite{Ghodsi:2013soa} where we focused on the point C in Fig.(\ref{fig1}.b), (Note that the $\lambda=1+\nu$ line is not within the parameter space of ring solutions, let alone the extremal rings). One may then wonder if the $\lambda=2,\ \nu=1$ point, which is also the end point of $\lambda=1+\nu$ line, can be obtained as a limit of the general case discussed earlier.

As we discussed in \cite{Ghodsi:2013soa}, a well-defined solution around $\lambda=2\,, \nu=1$ can be found by scaling the parameters as
\be \label{evh12}
\nu=1-\hat \nu \epsilon\,,\qquad \lambda=1+\nu-\hat\lambda \epsilon^{2(1+\alpha)}\,,\qquad k=\hat k \epsilon^{1+\alpha}\,.
\ee
With this scaling, the entropy and temperature behave as $S\sim \epsilon^{\alpha}$, $T\sim \epsilon^{2+\alpha}$, $J_{\phi}\sim \epsilon^{2\alpha}$ and the other charges remain finite. In other words we find that $T/S \sim \epsilon^2$ and moreover to get the EVH solution one should take $\alpha>0$.

The near horizon metric for the EVH ring can be easily obtained by inserting (\ref{evh12})\,, (\ref{trans1}) and (\ref{csca1}) with $(\alpha>1)$  and taking the limit $\epsilon \to 0$. The result will be a  metric like (\ref{nhDRevh}) solution with the AdS$_3$ radius $\ell^2=\frac{16\hat k^2}{\hat \lambda}$.

To find the near horizon of the near-EVH solution, it is enough to repeat all the steps similar to EVH case but with $\alpha=1$. In this case taking the limit $\epsilon \to 0$ yields the following metric
\be\begin{split}
ds^2=&\cos^2\theta \Big[-f(\hat r) d\hat t^2+\frac{d\hat r^2}{f(\hat r)}+\hat r^2\big( d\hat \phi-\frac{16\hat k^2}{\ell\,\hat r^2 \hat \nu^2}d\hat t \big)^2 \Big]+\ell^2\big(\cos^2\!\theta\, d\theta^2+\tan^2\!\theta d\hat \psi^2\big)\,,\\
f(\hat r)=& \frac{1}{\ell^2\hat r^2}\big(\hat r^2-\frac{16\hat k^2}{\hat \nu}\big)^2\,,\qquad \ell^2=\frac{16\hat k^2}{\hat \lambda}\,,
\end{split}\ee
which is a pinched extremal BTZ solution with radius $\ell^2=\frac{16\hat k^2}{\hat \lambda}$\,. The BTZ mass and spin also can be found easily as
\be \label{mjextbtz}
M_{BTZ}=\frac{r_+^2+r_-^2}{8\ell^2 G_3}=4\pi \hat k^2\,,\,\quad \qquad J_{BTZ}=\frac{r_+ r_-}{4\ell G_3}=4\pi \hat k^2 \ell\,.
\ee
The above expressions can be obtained as the $\nu\to 1$ limit of \eqref{mjbtz}. In other words, the  near horizon of \emph{extremal} near-EVH ring leads to an extremal BTZ, while for a generic near-EVH ring around  the $\lambda=1+\nu$ line we get a generic BTZ.

%%%%%%%%%%%%%%%%%%%%%%%%%%%%%%%%%%%%%%%%%%%%%%%%%%%%%%%%%%%%%%%%%%%%%%%%%%%%%%
\section{Unbalanced Pomeransky-Sen'kov black ring}\label{sec-3}

As discussed there are unbalanced ring solutions with conical singularity in the space-time \cite{{Emparan-Obers-2007},ur, Chen:2011jb}.
This is an asymptotically flat solution which contains the balanced Pomeransky-Sen'kov black ring as a special limit, where a ``balancing condition'' is satisfied.  A compact  form of this metric is \cite{Chen:2011jb}
\bea \label{UBR}
ds^2&=&-\frac{H(y,x)}{H(x,y)}\,\bigg(d t-\omega_\psi\,d\psi-\omega_\phi\,d\phi\bigg)^2-\frac{F(x,y)}{H(y,x)}\,d\psi^2-2\,\frac{J(x,y)}{H(y,x)}\,d\psi\,d\phi\nn\\ &+&\frac{F(y,x)}{H(y,x)}\,d\phi^2+\frac{2k^2(1-\mu)^2(1-\sigma)H(x,y)}{(1-\xi)(1-\mu\sigma)\Phi\Psi(x-y)^2}\bigg(\frac{d x^2}{G(x)}-\frac{d y^2}{G(y)}\bigg)\,,\nn\\
\omega_\psi&=&\frac{k(\mu+\sigma)}{H(y,x)}\,\sqrt{\frac{2\xi(\xi-\mu)(1+\xi)(1-\xi\mu)\Phi\Xi}{(1-\xi)(1-\mu\sigma)\Psi}}\,(1+y)\nn\\ &\times&\left\{\Phi ( 1+\sigma{x}^{2}y ) +\sigma(1-\mu)\left[1+\xi x -xy(x+\xi)\right]\right\},\nn\\
\omega_\phi&=&\frac{k(\mu+\sigma)}{H(y,x)}\,\sqrt{\frac{2\sigma\xi(1-\xi^2)\Phi\Psi\Xi}{1-\mu\sigma}}\,(1-x^2)y\,,\nn\\
\Phi\!&=&\!1-\!\xi\mu-\!\xi\sigma\!+\!\mu\sigma\,,\quad
\Psi\!=\!\mu-\!\xi\sigma+\!\mu\sigma-\! \xi\mu^2\,,\quad
\Xi\!=\!\mu+\!\xi\sigma-\mu\sigma-\!\xi\mu^2\,.
\eea
The functions $G$, $H$, $J$ and $F$ in the metric are given by
\bea
G(x)\!&=&\!(1-x^2)(1+\mu x)(1+\sigma x)\,, \nn\\ &&\cr
H(x,y)\!&=&\!\Phi\Psi+\sigma(\xi-\mu)(1+\xi)\Phi
+\sigma\Psi\Xi{x}^{2}{y}^{2}+\sigma ( \mu+\sigma )(\xi-\mu )(1- \xi\mu)( 1-\xi\mu{x}^{2}{y}^{2} ) \nn\\
&+&\! \xi ( \mu+\sigma)  ( 1-\xi\mu-\sigma (\xi-\mu)xy)((1-\xi\mu) x+ \sigma(\xi-\mu)y)\,,\nn\\ &&\cr
J(x,y)\!&=&\!\frac{2k^2(\mu+\sigma)\,\sqrt{\sigma(\xi-\mu)(1-\xi\mu)}(1-x^2)(1-y^2)}{(1-\mu\sigma)\Phi(x-y)}\,\Big\{
\Phi\Psi+\sigma(\xi-\mu)(1+\xi)\Phi \nn\\
&-&\! \sigma\Psi\Xi xy+\sigma
 ( \mu+\sigma )( \xi-\mu ) ( 1-\xi\mu ) ( 1+\xi x+\xi y+\xi\mu xy)\Big\}\,,\nn\\ &&\cr
F(x,y)\!&=&\!\frac{2k^2}{\mu\sigma(1-\mu\sigma)\Phi(x-y)^2}\,\bigg\{G(x)(y^2-1)\,\Big\{\mu(1-\xi^2)[\Psi+\sigma(\xi-\mu)(1+\sigma)]^2\cr
&-&\! (\mu+\sigma)(1-\xi\mu)(1+\sigma y)\big[\Psi\Xi-\xi\mu(\xi-\mu)[\Psi+\sigma(\xi-\mu)(1+\sigma)]\big]\Big\}\nn\\
&+&\!\sigma G(y)\Big\{(\xi-\mu)(1-\xi\mu)\big[
\xi(\mu+\sigma)^2(1-\xi\mu)+[\Psi+\sigma(\xi-\mu)(1+\sigma)]\nn\\
&\times&\! (\mu+\sigma-\mu\sigma x)x\big]+[\Psi\Xi+\xi\mu\Phi(\Phi-1)(\Phi-\Psi+\Xi)
][1+(\mu+\sigma)x]x^2\nn\\
&+&\! \mu\sigma\Phi[\Psi\Xi-\xi\mu(\mu+\sigma)(\xi-\mu)(1-\xi\mu)]x^4
\Big\}\bigg\}\,.
\eea
$x$ and $y$ coordinates are in the ranges $-1\leq x\leq 1$  and $-\infty<y< -1$ where infinity located at $x=y=-1$ and the azimuthal angles lie in the range $0\leq \phi,\psi \leq 2\pi$. Because of the unbalanced characteristic of the solution there are four independent parameters $\sigma, \mu, \xi, k$ instead of the three $\nu, \lambda, k$ in the Pomeransky-Sen'kov (balanced) metric.  The first three parameters are dimensionless while $k$ has dimension of length and determines the scale of the solution. The unbalanced ring parameter space is subject to
\be \label{ur ranges}
0\leq\sigma \leq \mu \leq \xi<1\,,\qquad k>0\,.
\ee

The metric (\ref{UBR}) has two horizons (roots of $G(y)$). The outer one is located at $y=-\frac{1}{\mu}$ and the inner  is at $y=-\frac{1}{\sigma}$\,.
The entropy, temperature, angular velocities, mass and spins of this solution are given by \cite{Chen:2011jb}
\bea \label{tur}
S\!&=&\!\frac{4\pi^2k^3(\mu+\sigma)(1-\mu)\Xi}{(1-\xi)(1+\mu)(1-\mu\sigma)^{3/2}}\,\left(\frac{2\xi(1+\xi)(1-\sigma)}{\Phi\Psi}\right)^{\!\frac12},\nn\\ T\!&=&\!\frac{(\mu-\sigma)(1-\xi)(1+\mu)}{8\pi k(\mu+\sigma)(1-\mu)\Xi}\,\left(\frac{2(1-\mu\sigma)\Phi\Psi}{\xi(1+\xi)(1-\sigma)}\right)^{\!\frac12},\nn\\
\Omega_\psi\!\!&=&\!\!\frac{1}{k(1\!-\!\mu)}\left(\frac{(\xi\!-\!\mu)(1\!-\!\xi)(1\!-\!\xi\mu)(1\!-\!\mu\sigma)\Psi}{2\xi(1+\xi)\Phi\Xi}\right)^{\!\!\frac 12},\quad \Omega_\phi\!=\!\frac{1+\mu}{k(\mu\!+\!\sigma)}\left(\frac{\sigma(1\!-\!\xi)(1\!-\!\mu\sigma)\Psi}{2\xi(1+\xi)\Phi\Xi}\right)^{\!\!\frac 12},\nn\\
M\!&=&\!\frac{3\pi k^2\xi(\mu+\sigma)(1-\mu)\Phi}{2(1-\xi)(1-\mu\sigma)\Psi}\,, \qquad
J_\phi=\frac{2\pi k^3(\mu+\sigma)(1-\mu)}{(1-\mu\sigma)^{3/2}}\,\left(\frac{2\sigma\xi(1+\xi)\Xi}{(1-\xi)\Phi\Psi}\right)^{\!\frac12}, \nn\\
J_\psi\!&=&\!\frac{\pi k^3(\mu\!+\!\sigma)(1\!-\!\mu)[2\sigma(1\!-\!\xi)(1\!-\!\mu)+(1\!-\!\sigma)\Phi]}{(1-\xi)^{3/2}(1-\mu\sigma)^{3/2}\Psi^{3/2}}\left(\frac{2\xi(\xi\!-\!\mu)(1\!+\!\xi)(1\!-\!\xi\mu)\Xi}{\Phi}\right)^{\!\frac 12}\,,
\eea
which satisfy the Smarr relation $TS-\Omega_{\phi} J_{\phi}-\Omega_{\psi} J_{\psi}-\frac23\,M=0$\,.

To determine existence of a  deficit angle in $\phi$  coordinate on the horizon one may expand the $\phi\phi$ component of the horizon metric around  $x=x_0$ where they vanish. The $x\!-\!\phi$ part of the metric takes the form
\be \label{defi1}
ds_H^2= A \frac{dx^2}{x-x_0}+B(x-x_0)d\phi^2\,,
\ee
where $x_0=\pm 1$. Using the transformation $x-x_0=\alpha r^2$, it is possible to rewrite this part as
\be\label{defi2}
ds_H^2=4\alpha A(dr^2+\kappa_E^2 r^2 d\phi^2)\,, \qquad \kappa_E^2=\frac{B}{4A}\,,
\ee
which exhibits a deficit or excess angle due to the periodicity $2\pi /\kappa_E$. For the metric (\ref{UBR}) at $x=1$ (the north pole of  $S^2$ at the horizon)
we find
\be \label{kappa}
\kappa_E=\frac{1+\mu}{1-\mu}\,\sqrt{\frac{(1-\xi)(1+\sigma)\Psi}{(1+\xi)(1-\sigma)\Xi}}\,,
\ee
while for $x=-1$ (the south pole of $S^2$ at the horizon) we find $\kappa_E=1$. This means that the unbalanced ring has a ``distorted $S^2$'' in its horizon which is a topologically $S^2$ geometry consisting of a conic space in its north pole joined to a round hemisphere at the south pole (a pear-shape geometry).
The conical singularity at the horizon of the unbalanced ring is removed if we put $\kappa_E=1$. This leads to a balancing condition between parameters as
\be \label{balance}
\xi=\frac{2\mu}{1+\mu^2}\,.
\ee
We note that value of $\psi \psi$ component on the horizon is also different at $x=\pm1$, but being a topologically $S^1$ direction, this means that  the radius of the ring varies on the horizon of unbalanced ring (ring is not geometrically a circle).

\subsection{The parameter space}
As it has been studied in \cite{Chen:2011jb}, the unbalanced ring in different limits contains  different solutions such as the  balanced Pomeransky-Senkov's black ring, the Emparan-Reall single rotating black ring \cite{Emparan:2001wn}, Figueras single rotating black ring \cite{Figueras:2005zp}, boosted Kerr string and the Myers-Perry black hole. In the following we will discuss the parameter space from a different point of view. Remembering the condition (\ref{ur ranges}), the parameter space is looking like a triangular pyramid, depicted in Fig(\ref{fig1}.a). Because of ranges in (\ref{ur ranges}), the $\xi=1$ face does not belong to the parameter space.  We  now concentrate on some special regions:
\begin{figure}[ht]
\begin{picture}(0,0)(0,0)
\put(100,-275){\footnotesize Fig (1.a)}
\put(345,-275){\footnotesize Fig (1.b)}
\end{picture}
%\vspace{-40mm}
\center
\includegraphics[height=165mm,angle=90]{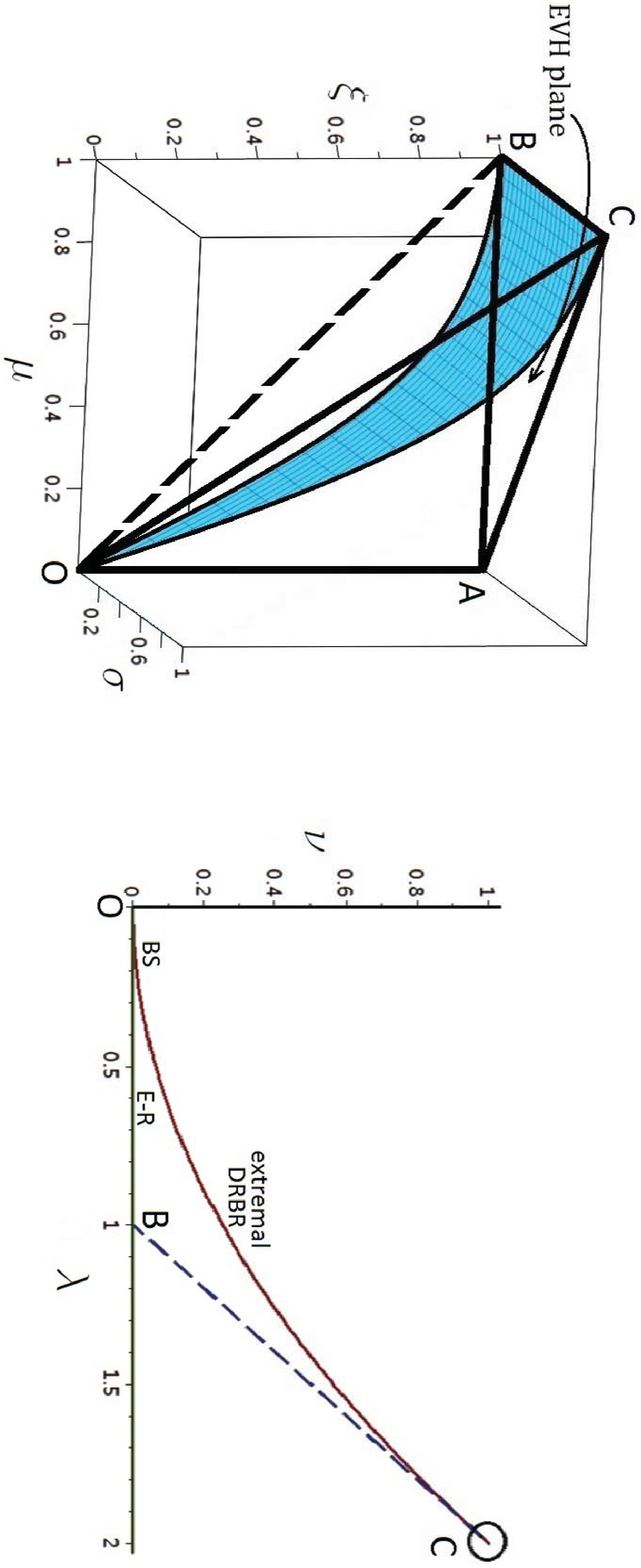}
\vspace{-14mm}
\caption{{\small Fig(1.a) The parameter space of the unbalanced black ring is located inside the (ABCO) pyramid. The $\xi=1$ face (the ABC triangle) does not belong to the parameter space. There are different known solutions which can be viewed as different limits of the unbalanced ring. The Pomeransky-Sen'kov (balanced) black rings are located on the blue curved plane surrounded by the pyramid; the  $\sigma=0$  or the (OAB) face represents the (unbalanced) Emparan-Reall single rotating black rings;  the $\xi=\mu$  or the (OBC) face recovers the Figueras black ring. On the (OB) line  both $J_{\phi}$ and $J_{\psi}$ vanishes and it is depicted by a dashed line. The extremal unbalanced rings lie on the $\mu=\sigma$ (OAC) face . In this case on the $\xi=1$ plane the solution becomes massless. The boosted Kerr strings are located around the origin (O), and the Myers-Perry black holes are around the $\xi=\mu=1$\, (BC) edge.
Fig(1.b) The parameter space of Pomeransky-Sen'kov (balanced) black ring. Points O, B and C in this figure are corresponding to the same points in Fig(1.a).
}\label{fig1}}
\end{figure}

\paragraph{$\bullet$ {\bf The balanced solution.}} By imposing the balance condition (\ref{balance}),  solution (\ref{UBR}) reduces to the Pomeransky-Sen'kov black ring with three parameters $(\lambda, \nu, k)$. To see this explicitly, one should use the relation between parameters of the balanced and unbalanced solutions as
\be \label{conv}
\lambda=\mu+\sigma\,, \qquad \nu= \mu\sigma\,,\qquad k=k\,.
\ee
We have shown the location of balanced solutions, computing from the balanced condition (\ref{balance}), with a blue curved plane in Fig(\ref{fig1}.a).
\paragraph{$\bullet$ {\bf The extremal face $\mu=\sigma$.}} Inserting $\mu=\sigma$ unifies the roots of $G(y)$ in (\ref{UBR}) and the temperature in (\ref{tur}) vanishes so this condition corresponds to the extremal unbalanced rings (OAC triangle in Fig (\ref{fig1}.a)).

\paragraph{$\bullet$ {\bf The  $\sigma, \mu, \xi \to 0$ corner.}} Upon the following transformations \cite{Chen:2011jb}
\bea
\sigma&=&\frac{m\!-\!\sqrt{m^2\!-\!a^2}}{\sqrt{2}k}\,, \quad \mu=\frac{m\!+\!\sqrt{m^2\!-\!a^2}}{\sqrt{2}k}\,, \quad \xi=\frac{m\!+\!\sqrt{m^2\!-\!a^2}}{\sqrt{2}k}\cosh\!\sigma\,,\nn \\ x&=&\cos\theta\,,\qquad \qquad y=-\frac{\sqrt{2}k}{r}\,,\qquad \qquad \psi=-\frac{z}{\sqrt{2}k}\,,
\eea
while taking the $k\to \infty$ limit keeping the new variables fixed, (\ref{UBR}) transforms to the boosted Kerr string.
The $k\to \infty$ limit takes $\sigma, \mu, \xi$ to zero. Therefore, the boosted Kerr string solution is obtained from the unbalanced ring around the origin O.
\paragraph{$\bullet$ {\bf The $\sigma=0$ face.}} Inserting $\sigma=0$ (OAB face), the solution (\ref{UBR}) recovers the known Emparan-Reall single spin black ring \cite{Emparan:2001wn}.
\paragraph{$\bullet$ {\bf The $\mu=\xi$ face.}} By considering $\mu=\xi$ the solution goes to the Figueras black ring \cite{Figueras:2005zp} which has only one rotation in the $\phi$ direction (the coordinate normal to the circle $S^1$ of the ring).
\paragraph{$\bullet$ {\bf The $\xi=\mu=1$ edge.}} Although the $\xi=1$ face does not belong to the parameter space, the solution shows interesting properties in the limit $\xi, \mu \to 1$. By inserting
\be \label{ximu}
\xi=1-c(1-\mu)\,,
\ee
where $0<c \leq 1$ and  the following coordinate transformations
\be \label{MPtrans}
x=-1\!+\!{\frac {8{k}^{2}  \cos^2\theta \left( 1-\mu \right) }{2{r}^{2}\!+a^{2}\!+b^{2}\!-m-4{k}^{2} \cos 2\theta}}\,, \quad
y=-1\!-\!{\frac {8{k}^{2} \sin^2\theta \left( 1-\mu \right) }{2{r}^{2}\!+a^{2}\!+b^{2}\!-m-4{k}^{2} \cos 2\theta}}\,,
\ee
and taking the $\mu\to 1$ limit, the solution (\ref{UBR}) goes over to the MP black hole. Here the mass and rotation parameters are given by
\be \label{mpc}
m\!=\!{\frac {4{k}^{2} \left( 1\!+\!\sigma \right) }{ c\left(1- \sigma \right)}}, \quad
a\!=\!{\frac { 2k\sqrt{(1\!-\!c^2)(1\!-\!\sigma)(1\!+\!\sigma\!+\!c\!-\!c\sigma)} }{\sqrt {c}\left( 1-\sigma+c+c\sigma \right) }},\quad
b\!=\!{\frac {4k\sqrt {c\sigma(1\!+\!\sigma\!+\!c\!-\!c\sigma)}}{\sqrt {1\!-\!\sigma}\left( 1\!-\!\sigma\!+\!c\!+\!c\sigma \right)}}\,.
\ee
Note that in the special case $c=1$, the rotation parameter $a$ is vanishing and solution reduces to the single rotating MP solution.

\subsection{The EVH limit}
In this section we  investigate whether the EVH condition (\ref{evhc}) can be satisfied in the parameter space of the unbalanced ring. We find that near  the (ABC) triangle in the upper plane $\xi=1$\,, the unbalanced ring solution becomes EVH. In fact by approaching  the $\xi=1$ plane as
\be \label{xiup}
\xi=1-\hat \xi \epsilon^a\,, \qquad k=\hat k \epsilon^{\frac{a}{2}}\,,
\ee
(scaling of $k$ is needed to keep the mass finite)  leads to the following expressions for the  charges of the solution
\bea \label{atup}
T&=&\frac{(1+\mu) \hat \xi \sqrt{(1-\mu\,\sigma)
(\mu-\sigma) ^{3}}}{8\pi(-1+\mu) (\mu+\sigma) ^{2}\hat k}\,{\epsilon}^{\frac a2}\,, \qquad
A_H = \frac {32{\pi }^{2}{\hat k}^{3} (\mu+\sigma)^{2}(-1+\mu )}{\sqrt{(\mu-\sigma)( 1-\mu\,\sigma)^{3}} \left( 1+\mu \right)\hat \xi }
\,{\epsilon}^{\frac a2}\,,\nn\\
M&=&\frac{3\pi \hat k^2(\mu+\sigma)(1-\mu)(1-\sigma)}{2\hat \xi(1-\mu \sigma)(\mu-\sigma)}+\mathcal{O}(\epsilon^a)\,,\qquad J_{\phi}=\frac{4\pi \hat k^3 (\mu+\sigma) \sqrt{\sigma(1-\mu)(\mu^2-\sigma^2)}}{(\mu-\sigma)\sqrt{\hat \xi (1-\sigma)(1-\mu \sigma)^3}}\epsilon^a\,,\nn\\
&&\qquad \qquad J_{\psi}=\frac{2\pi \hat k^3(1-\mu)^2(1-\sigma)^{3/2}(\mu+\sigma)^{3/2}}{\sqrt{\hat \xi^3(1-\mu)(\mu-\sigma)^3(1-\mu \sigma)^3}}+\mathcal{O}(\epsilon^a)\,,
\eea
$\Omega_{\phi}\,, \Omega_{\psi}$ also remain finite. If we denote the leading term in the expansion of charge $Q$ by $Q_0$ then we find the following relation
\be \label{btzsig}
M- \frac{M_0}{J_{\psi_0}}\,J_{\psi} \sim \epsilon^a\,.
\ee
As we will show this quantity can be related to the  mass of the BTZ solution which appears in the deviation from the EVH point.

In the  $\xi\to 1$ limit, to obtain the near horizon metric, one should take $\epsilon\to 0$ limit while inserting  (\ref{xiup}) and (\ref{csca1}) together with the following coordinate transformations\footnote{These transformations obtained by inserting $c=0$ in (\ref{MPtrans}) and (\ref{mpc}).}
\be \label{trans2}
x=-1\!+\!{\frac{8{k}^{2}\cos^2\theta (1-\mu)}{2{r}^{2}-4{k}^{2}\cos 2\theta}}, \qquad \quad
y=-1\!-\!{\frac{8{k}^{2}\sin^2\theta (1-\mu)}{2{r}^{2}-4{k}^{2}\cos 2\theta}}\,,
\ee
in the metric.
By choosing $a>2$ in (\ref{xiup}), we will approach to the EVH regime. Taking the limit $\epsilon \to 0$ the near horizon of the EVH unbalanced ring can be found as
\be
ds^2=\ell^2 \cos^2\theta\big(-\hat r^2 d\hat t^2+\frac{d\hat r^2}{\hat r^2}\big)+\frac{1-\sigma}{1-\mu \sigma}\,\hat r^2\cos^2\theta d\hat \phi^2+\ell^2\big(\cos^2\theta d\theta^2+\tan^2\theta d\hat \psi^2\big)\,,\nn
\ee
where $\ell^2=\frac{4\hat k^2(1-\sigma)(1-\mu)(\mu+\sigma)}{\hat \lambda(\mu-\sigma)(1-\mu \sigma)}$ and it can be related to the physical mass of the solution (\ref{atup}) as $M=\frac{3\pi \ell^2}{8}$. We can rewrite the near horizon metric by scaling $\hat t=\frac{\tilde t}{\ell^2}$ and $\hat \phi=\sqrt{\frac{1-\mu\sigma }{1-\sigma}}\tilde \phi$ into the usual pinching AdS$_3$ form
\be \label{abcnh}
ds^2=\cos^2\theta\Big[-\frac{\hat r^2}{\ell^2} d\tilde t^2+\ell^2 \frac{d\hat r^2}{\hat r^2}+\hat r^2 d\tilde \phi^2\Big]+\ell^2\big(\cos^2\theta d\theta^2+\tan^2\theta d\hat \psi^2\big)\,.
\ee
The  unbalanced characteristic of the original solution appears (only) through the range of pinched coordinate $\tilde \phi\in \Big[0,2\pi  \sqrt{\frac{1-\mu\sigma }{1-\sigma}}\epsilon\Big]$. Note that the AdS$_3$ radius and the value of pinching are different for each point on the (ABC) triangle.  When we are close to the (BC) line ($\mu=1$), which is the balanced limit, the range of the pinching coordinate reduces to $[0, 2\pi \epsilon]$, where the AdS$_3$ radius also  matches with the balanced case as we will see in the next subsection.

\subsubsection{EVH solutions on the BC line}
The BC line ($\xi \sim \mu \sim 1$) is the intersection of the balanced curved plane with the EVH solutions on ABC triangle, so one expects to find the balanced EVH solutions on it. In this subsection we study these solutions and their near horizons in the EVH and near EVH regimes. We will show the equivalence of the BC line in Fig.(\ref{fig1}.a) with the BC line in Fig.(\ref{fig1}.b).
In this limit to find a well-defined EVH solutions with finite charges, one should define the parameters as
\be \label{bcpar}
\xi=1-\hat \xi \epsilon^{2a}\,, \qquad \mu=1-\hat \mu \epsilon^a\,, \qquad k=\hat k \epsilon^{\frac a2}\,,
\ee
which leads to the following results for charges of the solution
\bea \label{bccharg}
T&=&\frac{\hat \xi (1-\sigma)^2}{4\pi \hat k \hat \mu (1+\sigma)^2}\,\epsilon^{\frac a2}, \qquad A=\frac{16\pi^2 \hat \mu \hat k^3(1\!+\!\nu)^2}{\hat \xi (1-\sigma)^2}\,\epsilon^{\frac a2},\qquad M=\frac{3\pi \hat \mu \hat k^2(1+\sigma)}{2\hat \xi (1-\sigma)}+\mathcal{O}(\epsilon^a)\,,\nn\\
J_{\phi}&=& \frac{4\pi \hat k^3 \sqrt{\hat \mu \sigma(1+\sigma)^3}}{\sqrt{\hat \xi (1-\sigma)^5}}\,\epsilon^a\,, \qquad \qquad J_{\psi}=\frac{2\pi \hat k^3 \hat \mu^{\frac 32}(1+\sigma)^{\frac 32}}{(1-\sigma)^{\frac 32} \hat \xi^{\frac 32}}+\mathcal{O}(\epsilon^a)\,.
\eea
Again we find the same behavior as (\ref{btzsig}) between the leading terms in the expansion of charges around the EVH point.
As in the previous sections, in order to find the near horizon we should insert (\ref{bcpar}), (\ref{csca1}) and (\ref{trans2}) into metric and take $\epsilon \to 0$ limit.

By considering $a>2$ we approach to the EVH regime and the near horizon metric takes the following form
\be \label{bcevhnh}
ds^2=\ell^2 \cos^2\theta\Big[-\hat r^2 d\hat t^2+\frac{d\hat r^2}{\hat r^2}+\frac{\hat r^2}{\ell^2} d\hat \phi^2\Big]+\ell^2\big(\cos^2\theta d\theta^2+\tan^2\theta d\hat \psi^2\big)\,,
\ee
where the AdS$_3$ radius is $\ell^2=\frac{8M}{3\pi}=\frac{4\hat k^2 \hat \mu(1+\sigma)}{\hat \xi (1-\sigma)}$ and the period of $\hat \phi$ is reduced to $2\pi \epsilon$, so (\ref{bcevhnh}) is a balanced solution. After another rescaling $\tilde t=\ell^2 \hat t$ one can find a pinching AdS$_3$ geometry as
\be \label{bcnh}
ds^2=\cos^2\theta\Big[-\frac{\hat r^2}{\ell^2} d\tilde t^2+\ell^2 \frac{d\hat r^2}{\hat r^2}+\hat r^2 d\hat \phi^2\Big]+\ell^2\big(\cos^2\theta d\theta^2+\tan^2\theta d\hat \psi^2\big)\,.
\ee

If we consider $a=2$ we will  then approach to the near-EVH regime. Note also that (\ref{btzsig}) will be proportional to $\epsilon^2$ which is a signature of a BTZ solution. In this case, upon the redefinition $\hat r=\frac{\sqrt{\rho^2(\sigma-1)^2-2\hat k(\sigma^2+6\sigma+1)}}{1-\sigma}$, the near horizon metric takes to the following form
\be\label{bcnevhnh}\begin{split}
ds^2=&\cos^2\theta \Big[-\!{f(\rho)}d\hat t^2+\frac{d\rho^2}{f(\rho)}+\rho^2 \big(d\hat\phi^2+\frac{8\hat k^2\sqrt{\sigma}(1+\sigma)}{\rho^2\ell\,(\sigma-1)^2}d\hat t\big)^2\Big]\!+\!\ell^2\big(\cos^2\theta d\theta^2+\tan^2\theta d\hat \psi^2\big)\,,\\
f(\rho)=& \frac{1}{\ell^2\rho^2}\Big(\rho^2-\frac{4\hat k^2(\sigma+1)^2}{(\sigma-1)^2}\Big)\Big(\rho^2-\frac{16\hat k^2\sigma}{(\sigma-1)^2}\Big)\,,\qquad \ell^2=\frac{4\hat k^2 \hat \mu(1+\sigma)}{\hat \xi(1-\sigma)}\,,
\end{split}\ee
which is a (pinching) BTZ with the mass and spin
\be \label{mjubtz}
M_{BTZ}=\frac{\pi \hat k^2 (\sigma^2+6\sigma+1)}{2(1-\sigma)^2}\,, \qquad
J_{BTZ}=\frac{4\pi \hat k^3 \sqrt{\hat \mu \sigma(1+\sigma)^3}}{\sqrt{\hat \xi (1-\sigma)^5}}=J_{\phi_0}\,.
\ee
We can see the agreement of these charges with the mass and spin (\ref{mjbtz}) on the BC line in the parameter space of the balanced ring in Fig.(\ref{fig1}.b).
Using (\ref{conv}) and (\ref{bcpar}) the balanced condition takes the form of $\hat \xi=\hat \mu^2/2$. It can be easily checked that the BTZ masses are exactly equal and the BTZ spins are equal when $\hat \lambda=\hat \mu(1-\nu)$.
We can verify this relation in a geometrical approach: Using the relations (\ref{bcpar}) or (\ref{parscane}) we approach to the BC line from different directions but $\hat \lambda=\hat \mu(1-\nu)$, unifies both directions into one.
\subsubsection{The extremal line AC}
As mentioned, the extremal unbalanced ring is located on the $\mu=\sigma$ plane in the parameter space. On the other hand the AC line is the intersection of the extremal plane with the EVH face $\xi=1$, so we can study the EVH solutions in this limit.
To find a well-defined solution one should scale the parameters as follows
\be \label{acpar}
\mu=\sigma+ \hat \mu \epsilon^a\,, \qquad \xi=1-\hat \xi \epsilon^{\frac a2}\,, \qquad k=\hat k \epsilon^{\frac a2}\,.
\ee
Again the near horizon limit can be found by inserting (\ref{acpar}) and (\ref{csca1}) together with (\ref{trans2}) and taking the  $\epsilon \to 0$ limit. For EVH solutions ($a>2$), the near horizon metric can be found by an additional rescaling $\tilde t=\frac{\hat t}{\sqrt{1+\sigma}}$ as
\be \label{acevhnh}
ds^2=\cos^2\theta\Big[-\frac{\hat r^2}{\ell^2} d\tilde t^2+\ell^2 \frac{d\hat r^2}{\hat r^2}+\hat r^2 d\tilde \phi^2\Big]+\ell^2\big(\cos^2\theta d\theta^2+\tan^2\theta d\hat \psi^2\big)\,,
\ee
with $\ell^2=\frac{8\hat k^2(\sigma-1)^2}{\hat \xi^2 (\sigma+1)^2}$\,. Due to the range of pinched coordinate  $\tilde \phi \in \left[0, \frac{2\pi}{\sqrt{1+\sigma}}\,\epsilon\right]$, the above metric is describing an unbalanced pinched AdS$_3$ geometry.

%One can discuss about a dual 2D CFT for such a solution \cite{BH}, with a central charge equal to $c=\frac{3\ell}{2G_3}$. We postpone this to the next sections.
%%%%%%%%%%%%%%%%%%%%%%%%%%%%%%%%%%%%%%%%%%%%%%%%%%%%%%%%%%%%%%%%%%%%%%%%%%%%%%
\section{Double rotating dipole ring}\label{sec-4}
A generalization of the double rotating black ring \cite{Pomeransky:2006bd} containing dipole charges was introduced in \cite{Chen:2012kd} (see also \cite{dipole, Yazadjiev}). This is a solution of Einstein-Maxwell-dilaton theory
\be
I =\frac{1}{16\pi G_5} \int d^5 x \sqrt{-g}\,\big[R-\frac 12\, \partial_\mu\varphi\partial^\mu\varphi-\frac 14\, e^{-\alpha\varphi}F_{\mu\nu}F^{\mu\nu}\big]\,.
\ee
For particular values of parameter $\alpha$, $\alpha^2=4/N-4/3,\ N=1,2,3$ the above action may be embedded in string theory compactification \cite{dipole}. For the  $N=1, \alpha=\sqrt{8/3}$ case the dipole ring solution is \cite{Chen:2012kd}
\bea \label{drdr}
ds^2_5&=&-\left[\frac{H(y,x)^3}{K(x,y)^2H(x,y)}\right]^{1/3}\left(d t+\omega_1\,d\psi+\omega_2\,d\phi\right)^2+\frac{2R^2}{(x-y)^2}\left[K(x,y)H(x,y)^2\right]^{1/3}\nn \\
&\times&\bigg\{\frac{F(x,y)\left(d\psi+\omega_3\,d\phi\right)^2}{H(x,y)H(y,x)}-\frac{G(x)G(y)\,d\phi^2}{F(x,y)}+\frac{1}{\Phi\Psi}\bigg[\frac{d x^2}{G(x)}-\frac{d y^2}{G(y)}\bigg]\bigg\}\,.
\eea
Coordinates lie in the ranges $0\leq \phi, \psi\leq 2\pi\,, \, -1\leq x \leq 1$ and $-\infty <y < -1$\,; also functions of the metric are as followings
\be \label{funcs}
\begin{split}
G(x)&=(1-x^2)(1+cx)\,,\\
K(x,y)&=-a^2(1+b)\big[bx^2(1+cy)^2+(c+x)^2\big]+\big[b(1+cy)-1-cx\big]^2+bc^2(1-xy)^2, \\
H(x,y)&=-a^2(1+b)\left[b(1+cx)(1+cy)xy+(c+x)(c+y)\right]-a(1+b)(x-y)\big[c^2-1 \\
&+b(1+cx)(1+cy)\big]+\left[b(1+cy)-1-cx\right]\left[b(1+cx)-1-cy\right]+bc^2(1-xy)^2, \\
F(x,y)&=\frac{1-y^2}{\Phi\Psi}\,\Big(bcG(x)\Big\{c(y^2-1)\left[a^2(1+b)-b+1\right]^2-4a^2y(1-b^2)(1+cy)\Big\} \\
&-(1+cy)\Big\{a^2(1+b)^2\left[a^2(c+x+bx+bcx^2)^2-(c+x-bx-bcx^2)^2\right] \\
&-(1-b)^2(1+cx)^2\left[a^2(1+b)^2-(1-b)^2\right]\Big\}\Big)\,, \\
J_{\pm}(x,y)&=a^2(1+b)\left[bx(1+cx)(1+cy)+(1+c)(c+x)\right]-bc^2(1-x)(1-xy) \\
&\pm a\left\{(1-x)\left[b(1+cx)+c-1\right]\left[b(1+cy)+c+1\right]-2bc(1-y)(1+cx)\right\} \\
&-\left[b(1+cx)-c-1\right]\left[b(1+cy)-cx-1\right]\,,\\
L(x,y)&=a^2(1+b)\big[bx(1+cy)^2+(1+c)(c+x)\big]-a(1-x)\big[b^2(1+cy)^2+c^2-1\big]\\
&-\big[b(1+cy)-c-1\big]\big[b(1+cy)-cx-1\big]-bc^2(1-y)(1-xy)\,,
\end{split}
\ee
together with
\be\begin{split}
\omega_1&=\sqrt{\frac{2a(a+c)}{\Phi\Psi}}\,\frac{R(1+b)(1+y)J_{+}(x,y)}{H(y,x)}\,,\\
\omega_2&=\sqrt{\frac{2ab(a+c)(1-a^2)}{\Phi\Psi}}\,\frac{R c(1+b)(1-x^2)\big[(1+cy)(a+ab+by)-c-y\big]}{H(y,x)}\,,\\
\omega_3&=\frac{\sqrt{b(1-a^2)}}{\Phi\Psi}\,\frac{ac(1+b)(x-y)(1-x^2)(1-y^2)}{F(x,y)}\\
&\times\left[b(1+cx)(1+cy)(1-b-a^2-a^2b)-(1-c^2)(1-b+a^2+a^2b)\right]\,,
\end{split}
\ee
where
\be
\Phi=1+a-b+ab, \qquad \Psi=1-a-b-ab.
\ee
This solution has four independent parameters $a, b, c$ and $R$. The parameter $a$ controls the dipole charge, $b$ introduces the rotation on the $S^2$, size of the black ring is characterizes by $c$ and $R$ represents the scale of the solution. These parameters are satisfying the following constraints
\be \label{pranges}
0\leq c\leq a<1\,,\qquad 0\leq b< \frac{1-a}{1+a}\,,\qquad R>0\,.
\ee
The gauge field $A$ and dilaton field $\varphi$ are given by
\be\label{dilg}
A=A_t dt+A_{\phi} d\phi+A_{\psi} d\psi\,, \qquad e^{-\varphi}=\left(\frac{K(x,y)}{H(x,y)}\right)^{\sqrt{2/3}}\,,
\ee
where
\be\label{gfs}
\begin{split}
A_{t}=&-\sqrt{b(a^2-c^2)(1-a^2)}\ \frac{c(1+b)(1-xy)(x-y)}{K(x,y)}\,,\\
A_{\phi}=&-\sqrt{\frac{2a(a-c)}{\Phi\Psi}}\ \frac{R(1+b)(1+x)L(x,y)}{K(x,y)}\,,\\
A_{\psi}=&-\sqrt{\frac{2ab(a\!-\!c)(1\!-\!a^2)}{\Phi\Psi}}\,\frac{R c(1\!+\!b)(1\!+\!y)}{K(x,y)}\big(x(1\!-\!y)(1\!+\!c)\Phi+(1\!-\!x)^2(a+ab+bcy+c)\big)\,.
\end{split}
\ee
The metric (\ref{drdr}) is an asymptotically flat solution with infinity located at $x=-1, \, y=-1$ \cite{Chen:2012kd}. The ADM mass and angular momenta of the solution are given by
\be\label{mj}
\begin{split}
M&=\frac{\pi R^2(1+b)[(a+c)\Phi+a(1-b+c+bc)]}{\Phi\Psi}\,,\\
J_\psi&=\frac{2\pi R^3(1+b)[(1+c)\Phi+2bc(1-a)]}{\Psi^{3/2}}\,\sqrt{\frac{a(a+c)}{2\Phi}}\,,\\
J_\phi&=\frac{2\pi R^3c(1+b)}{\Psi^{3/2}}\,\sqrt{\frac{2ab(a+c)(1-a^2)}{\Phi}}\,.
\end{split}\ee
The entropy, temperature, horizon angular velocities, magnetic dipole charge and potential are given by
\bea \label{termdrd}
S&=&4\pi^2R^3 c(1+b)\sqrt{\frac{2a(a+c)(1-a^2)}{\Phi\Psi^3}}\,,\qquad \quad \,\,\,T=\frac{1}{8\pi (1+b)R}\,\sqrt{\frac{2\Phi\Psi^3}{a(a+c)(1-a^2)}}\,,\nn\\
\Omega_\phi&=&\frac{1\!-\!b\!+\!a^2\!+a^2b}{R\,(1+b)}\,\sqrt{\frac{b\Psi}{2a(a\!+\!c)(1\!-\!a^2)\Phi}}\,, \quad \quad \,\,\, \Omega_\psi=\frac{1}{R}\,\sqrt{\frac{a\Psi}{2(a+c)\Phi}}\,,\nn\\
Q&=&\frac{(1+b)R\sqrt{2a(a-c)}}{\sqrt{\Psi\Phi}}\,, \qquad \qquad \qquad\,\,\, \qquad \Phi_m=\frac{\pi R \sqrt{2(a-c)\Psi} }{2\sqrt{a\Phi}}\,,
\eea
which satisfy the Smarr formula $\frac 23 M=TS+\Omega_{\phi} J_{\phi}+\Omega_{\psi} J_{\psi}+\frac{1}{3} Q \Phi_m$ and the first law of thermodynamics
with the dipole charge and its potential also included \cite{Chen:2012kd}.
%%%%%%%%%%%%%%%%%%%%%%%%%%%%%
\subsection{Parameter space and extremal limit}
\begin{figure}[tt]
\vspace{-20mm}
\center
\includegraphics[height=150mm,angle=0]{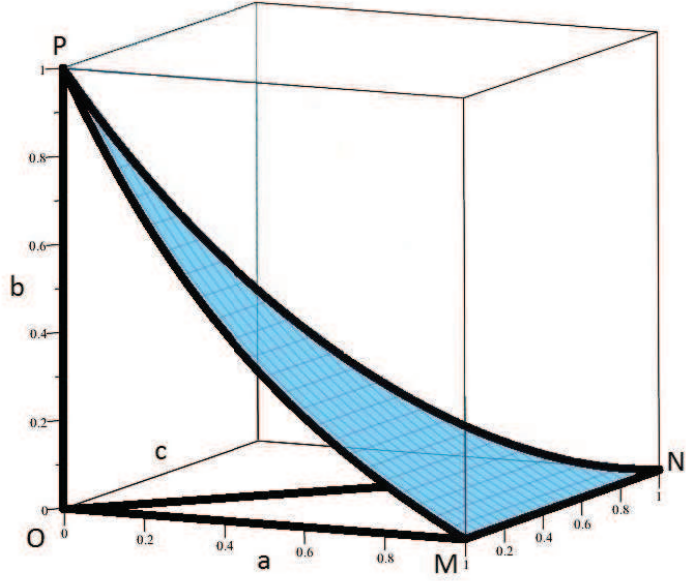}
\vspace{-34mm}
\caption{{\small The parameter space of the dipole rings. All double rotating dipole rings are inside the deformed pyramid (OMNP). Single rotating dipole rings are located on the OMN face. The ONP face is the location of the neutral double rotating rings. This face is equivalent to the blue plane OBC in Fig.(\ref{fig1}.a). The extremal double rotating MP black holes lie near the MNP plane. The EVH condition is satisfied for black rings/holes on the MN edge. This edge is equal to the BC edge in Fig.(\ref{fig1}.a}).
\label{fig2}}
\end{figure}

Under the conditions in (\ref{pranges}), the parameter space of the solutions, is a ``pyramid like'' space (OMNP) as depicted in Fig.(\ref{fig2}). Note that the blue face (MNP) does not belong to the parameter space but we can study its neighborhood. All double rotating dipole rings lie inside this pyramid. Also the neutral double rotating black rings, single rotating dipole rings and MP black holes can be studied as different limits of this solution.

\paragraph{$\bullet$ {\bf The ONP face.}}
As  (\ref{gfs}) shows, by inserting $a=c$ one eliminates the dipole charge so the ONP face represents the neutral double rotating black rings \cite{Pomeransky:2006bd}.

\paragraph{$\bullet$ {\bf The OMN face.}}
The parameter $b$ tunes the rotation on $S^2$. By choosing $b=0$, two parameters $\omega_2$ and $\omega_3$ \eqref{funcs} become zero and this turns off rotation in the $\phi$ direction. In other words, the single rotating dipole rings \cite{dipole} are located at $b=0$ plane.

\paragraph{$\bullet$ {\bf The MNP plane.}} At this surface $\Psi=0$ or by $b=\frac{1-a}{1+a}$. As one can see from \eqref{mj}, in order to keep the mass and spins finite one is forced to take $R$ to zero. This is hence a collapsing ring limit where the ring collapses into a black hole. This is somehow similar to the BC line in Fig.(\ref{fig1}.b). Finite mass can be guaranteed if we take $R\to 0$ limit introducing a new parameter $\eta$ defined through
\be \label{collaps}
b=\frac{1-a-\eta R^2}{1+a}\,.
\ee
Defining the following parameters
\be \label{d2rmp}
m=\frac{4(a+c)}{\eta(1+a)}\,,\qquad a_1=\frac{\sqrt{2}(a^2c+2a+c)}{\sqrt{\eta(1+a)(a+c)}}\,,\qquad a_2=\frac{\sqrt{2}c(1-a^2)}{\sqrt{\eta(1+a)(a+c)}}\,,
\ee
and performing the following coordinate transformations  \cite{Chen:2012kd}
\be \label{d2rct}
x=-1+\frac{4R^2(1-c)\cos^2\theta}{r^2-a_1a_2}\,,\qquad y=-1-\frac{4R^2(1-c)\sin^2\theta}{r^2-a_1a_2}\,,
\ee
one can explicitly find the 5D MP black hole rotating in both $\phi$ and $\psi$ directions
\bea \label{MPbh}
ds^2&=& -\frac {\Delta}{{\rho}^{2}} \left[ {dt}-a_1 \sin^2 \theta d\phi -a_2 \cos^2 \theta d\psi \right] ^{2}+\rho^2 \big({\frac {{dr}^{2}}{\Delta}}+d\theta^2\big) \nn\\
&+& \frac {\sin^2 \theta}{{\rho}^{2}} \big[ a_1{dt}- \left( {r}^{2}+{a_1}^{2}
 \right) d\phi  \bigr] ^{2} +{\frac { \cos^2 \theta}{{\rho}^{2}} \left[ a_2{dt}- \left( {r}^{2}+
{a_2}^{2} \right) d\psi  \right] ^{2}}\nn \\
&+&{\frac {1}{{r}^{2}{\rho}^{2}} \left[ a_1a_2\,{dt}-a_2 \left( {r}^{2}+{a_1}^{2} \right) \sin^2 \theta d\phi -a_1 \left( {r}^{2}+{a_2}^{2} \right) \cos^2 \theta d\psi  \right] ^{2}}\,,
\eea
where $\Delta=\frac{1}{r^2}(r^2 + a_1^2)(r^2 + a_2^2) - 2m$\,, and $\rho^2 = r^2 + a_1^2\cos^2\theta + a_2^2\sin^2\theta\,.$

As we will see below, in the collapsing limit the temperature of the black ring tends to zero and it becomes extremal. The resulting MP black hole, too,  will have a vanishing temperature, as one can readily check from \eqref{d2rmp}, the parameters $a_1$ and $a_2$ of the MP black hole satisfy extremality condition $m=\frac12(a_1+a_2)^2$.

\paragraph{$\bullet$ {\bf The extremal  corner P.}}
In addition to the collapsing limit $R\to 0$\,, there is another way to get an extremal solution. If we settle in the region $a,c \to 0$ and $b \to 1$ in the parameter space, then the temperature will vanish and the geometry of (\ref{drdr}) becomes extremal.

%%%%%%%%%%%%%%%%%%%%%%%%%%%%%%%%%%%%%%%%%%%%%%%%%%%%%%%%
\subsection{The EVH limit}
As in the other black ring solutions studied here and in \cite{Ghodsi:2013soa}, the EVH condition is satisfied in the collapsing limit. Using (\ref{collaps}), defining $R = \hat R\,\epsilon^{\delta}, \, a=1-\epsilon^{\gamma}$
and taking the limit $\epsilon \to 0$, the temperature and entropy in (\ref{mj}) of the solution behave as $T\sim \hat R^2 \epsilon^{2\delta-\gamma/2}$ and $S\sim \epsilon^{\gamma/2}$.
In fact the double rotating dipole ring becomes EVH if $0<\gamma<4\delta\,$ but for simplicity we will consider $\gamma=\delta$
\be \label{drdrevh}
b=\frac{1-a-\eta R^2}{1+a}\,, \qquad R = \hat R\,\epsilon^\delta\,, \qquad a=1-\epsilon^{\delta}\,.
\ee
In the EVH point the entropy, temperature and charges of the solution are as follows
\bea
S&=&\frac{4\pi^2 c \sqrt{2 \hat a(1+c)}}{\eta^{\frac 32}}\,\epsilon^{\frac{\delta}{2} }\,, \qquad T=\frac{\sqrt{2\eta^{3}}\,\hat R^2}{4\sqrt{\hat a(1+c)}}\,\epsilon^{\frac{3\delta}{2}}\,, \qquad M=\frac{3\pi (1+c)}{2\eta}+\mathcal{O}(\epsilon^{\delta})\,,\nn\\
J_{\phi}&=&\frac{2\pi c \hat a \sqrt{1+c}}{\eta^{3/2}}\,\epsilon^{\delta}\,, \qquad J_{\psi}=\frac{2\pi (1+c)^{3/2}}{\eta^{3/2}}+\mathcal{O}(\epsilon^\delta)\,,
\eea
and angular velocities also remain non-vanishing. Note that in the collapsing limit, even without sending $a\to 1$\,, the black ring solution will be extremal. In fact for the MP solution, the parameters in (\ref{d2rmp}) satisfy the extremality condition $m=\frac12(a_1+a_2)^2$ and when we send $a\to 1$ the MP black hole becomes single rotating, and hence EVH MP black hole \cite{Ghodsi:2013soa}.

Another interesting fact is that in the EVH limit, the gauge field components (\ref{gfs}) vanish  and the resulting solution becomes uncharged
\bea
A_t&=&\frac{8\hat a c \sqrt{1-c^2}\,\hat R^2(1-2\cos^2\!\theta)}{4r^2(1+c)\cos^2\!\theta+r^4 \eta}\,\epsilon^6, \qquad A_{\phi}=\frac{4\hat a c \sqrt{1-c}\,\hat R^2\cos^2\!\theta}{r^2 \sqrt{\eta}}\,\epsilon^6\,, \nn\\
A_{\psi}&=&-\frac{4c\hat a \sqrt{1-c}\,\hat R^2\sin^2\!\theta\big[r^2\eta+4(1+c) \sin^2\theta \big]}{r^2 \sqrt{\eta}\,\big[r^2\eta+4(1+c) \cos^2\theta\big]}\,\epsilon^6\,.
\eea
This has a simple physical explanation: The EVH condition occurs when $(x,y)$ coordinates are computed around the $(-1,-1)$ point, which  corresponds to the asymptotic region of the ring.
On the other hand the dipole ring is a distribution of the magnetic monopoles \cite{dipole} and
the electromagnetic fields of the dipole ring outside the horizon is produced by a circular array of magnetic monopoles
such that, despite of having  a local distribution of charge, the total magnetic charge is zero.
Therefore, the dipole ring  can not be distinguished from a neutral ring from the asymptotic infinity and electromagnetic fields are vanishing in the EVH limit.

In order to discuss the near horizon of  EVH solutions, one should insert (\ref{drdrevh}) and (\ref{csca1}) together with (\ref{d2rct}) in the original solution (\ref{drdr}) and take the  $\epsilon \to 0$ limit.
Considering $\delta>2$, one can settle in the EVH regime. In this case by an extra rescaling $\tilde t=\ell^2 \hat t$, the near horizon metric takes the following form
\be \label{colevh}
ds^2=\cos^2\theta\Big[-\frac{\hat r^2}{\ell^2} d\tilde t^2+\ell^2 \frac{d\hat r^2}{\hat r^2}+\hat r^2 d\hat \phi^2\Big]+\ell^2\big(\cos^2\theta d\theta^2+\tan^2\theta d\hat \psi^2\big)\,,
\ee
which is the familiar pinched  AdS$_3$ with radius $\ell^2=\frac{8M}{3\pi}=\frac{4(1+c)}{\eta}$.

The near-EVH solution is obtained for $\delta=2$. In this case we find that the near horizon of the near-EVH ring takes the form of \emph{extremal} pinching BTZ solution
\be\begin{split}
ds^2=&\cos^2\theta\ \Big[- f(\hat r) d\hat t^2+\frac{d\hat r^2}{f(\hat r)}+\hat r^2\big(d\hat \phi-\frac{4\eta\, \hat a c}{\ell r^2}\, d\hat t \big)^2\Big]+ \ell^2\big(\cos^2\theta d\theta^2+\tan^2\theta d\hat \psi^2\big)\,,\\
f(\hat r)=& \frac{\big(\hat r^2-\frac{4\hat a c}{\eta} \big)^2}{\ell^2 \hat r^2}\,,\qquad \ell^2=\frac{4(1+c)}{\eta}\,.
\end{split}\ee
%%%%%%%%%%%%%%%%%%%%%%%%%%%%%%%%%%%%%%%%%%%%%%%%%%%%%%%%%%%%%%%%%%%%%%%
\section{The EVH/CFT and black rings}\label{sec-5}

As we discussed in the near horizon geometry of EVH  black rings we find a (pinching) AdS$_3$ throat which turns into a BTZ for near-EVH solutions.
One may then use this AdS$_3$ to extend the EVH/CFT proposal \cite{SheikhJabbaria:2011gc} for these rings too. The dual 2d CFT proposed here would hence govern the low energy excitations or perturbations around the original EVH black ring.
Below we briefly discuss this dual CFT for different EVH rings we discussed in previous sections.

\subsection{The unbalanced ring}

As we saw in the previous sections for unbalanced ring the EVH conditions are satisfied  at $\xi=1$. Besides the EVH point, the unbalanced ring becomes extremal
at $\sigma=\mu$, where as we will discuss there is also a Kerr/CFT type chiral 2d CFT \cite{Guica:2008mu,{Chen:2012yd}} to these geometries.

\subsubsection{The EVH solution near $\xi=1$, (ABC triangle)}
In the previous sections we showed the explicit form of the near horizon metrics in the EVH limits (\ref{abcnh}). Since this metric contains an AdS$_3$ part we can find the Brown-Hennueax \cite{BH} central charge $c_{B.H.}=\frac{3 \ell}{2G_3}$ of the 2d CFT associated with this AdS$_3$ factor. To find the 3d Newton constant $G_3$ in terms of the $G_5$ we use the reduction ansatz
\bea
ds^2= \cos^2 \theta \, \underbrace{g_{\alpha \beta} dx^{\alpha} dx^{\beta}}_{AdS_3}+\ell^2 \big(\cos^2\theta d \theta^2+\tan^2 \theta d \psi^2 \big)\,,
\eea
where by considering $G_5=1$ the result is
\be \label{grel}
G_3=\frac{1}{\pi \ell^2}\,.
\ee
One should note that due to the pinching the above Brown-Henneaux central charge is for the 2d CFT on a pinching orbifold cylinder, which can be equivalent to a 2d CFT at central charge $c_{B.H.}\epsilon$ on a cylinder (without pinching) \cite{deBoer:2010ac}.

\paragraph{The region inside the ABC triangle.}
 The Brown-Hennueax central charge in this case can be read from the near horizon metric (\ref{abcnh})
\be
c_{B.H.}=\frac{3\ell}{2G_3}=\frac{3\pi \ell^3}{2}=12\pi \hat k^3 \left[\frac{(1-\sigma)(1-\mu)(\mu+\sigma)}{\hat \lambda(\mu-\sigma)(1-\mu \sigma)}\right]^{\frac 32}\,.
\ee
\paragraph{The BC line ($\mu =\xi= 1$ edge).}
We can find the Brown-Hennueax central charge for this case from (\ref{bcnh}) as
\be \label{ccbc}
c_{B.H.}=\frac{3\ell}{2G_3}=\frac{3\pi \ell^3}{2}=12\pi \hat k^3 \left[\frac{\hat \mu(1+\sigma)}{\hat \xi (1-\sigma)}\right]^{\frac 32}\,.
\ee
On the other hand on $\xi=1$ plane the value of $c$ in  (\ref{ximu}) vanishes and hence the rotation parameter  $b$ in (\ref{mpc}) is also vanishing. To keep the mass $m$ and rotation parameter $a$ finite, we need to rescale $k$ and $c$ in (\ref{mpc}) as $k=\hat k \,\epsilon, c=\hat c \,\epsilon^2$ and the geometry becomes that of a  single rotating extremal MP black hole with $2m=a^2=\frac{4\hat k^2(1+\sigma)}{1-\sigma}$. It was discussed in  \cite{Ghodsi:2013soa} that  the central charge for this EVH MP black hole is given by $c_{CFT}=\frac{3\pi}{2}\,a^3=12\pi \hat k^3\left[\frac{1+\sigma}{\hat c(1-\sigma)}\right]^{3/2}$. Recalling (\ref{ximu}) and the definitions of $\xi$ and $\mu$ in (\ref{bcpar}), this central charge as expected matches with the value in (\ref{ccbc}).

\paragraph{The AC line $\sigma=\mu\,; \,\xi= 1$.}
Noting (\ref{acevhnh}) and that $\ell^2=\frac{8\hat k^2(\sigma-1)^2}{\hat \xi^2 (\sigma+1)^2}$, it is possible to read  the central charge as
\be
c_{B.H.}=24\sqrt{2}\,\pi\, \hat k^3\frac{(\sigma-1)^3}{\hat \xi^3 (\sigma+1)^3}\,.
\ee
\subsubsection{The extremal plane $\sigma=\mu$}

For completeness, here we briefly  discuss the extremal, but non-EVH, (unbalanced) ring and its chiral dual 2d CFT via Kerr/CFT \cite{Guica:2008mu,Chen:2012yd}. To this end, we need to read the near horizon metric of the extremal ring:
\be
ds^2=\alpha(x) \big(-y^2 dt^2+\frac{dy^2}{y^2}\big)+\beta(x)dx^2 +\gamma\, d\psi^2+\delta(x)(d\phi+\rho \, d\psi+f^{\phi} y dt)\,,
\ee
where $f^{\phi}=1$ and the other functions and parameters are given by
\bea
\alpha (x) &=&-{\frac {2{\mu}^{2} ( \mu-1 ) ^{2}{k}^{2} ( 1+\xi )  ( {x}^{2}+2 \xi x+1
 ) }{ ( 1-2 \xi \mu+{\mu}^{2} )  ( 1+\mu x ) ^{2} ( \xi-1 )  ( 1+\mu ) ^{2}}}\,,\nn \\
\beta(x)&=&-\frac {2{k}^{2} ( \mu-1 ) ^{3}{\mu}^{2} ( 1+\xi )  ( {x}^{2}+2 \xi x+1
 ) }{ ( 1-2 \xi \mu+{\mu}^{2} )  ( 1+\mu x ) ^{4} ( 1+\mu )  ( \xi-1 )  (
x-1 )  ( x+1 ) } \,,\nn \\
\delta(x)&=&-\frac {8{k}^{2}{\mu}^{2} ( -1+{\xi}^{2} )  ( {x}^{2}-1 ) }{ ( -1+{\mu}^{2} )  ( {x}^{2}+2 \xi x+1 )  ( 1-2 \xi \mu+{\mu}^{2} ) } \,,\nn \\
\gamma&=&-\frac {2{k}^{2}\lambda\, \left( 1+\lambda \right)  \left( \mu-1
 \right) ^{2} \left( 1-2\,\lambda\,\mu+{\mu}^{2} \right) }{ \left(
\lambda-1 \right) ^{3} \left( 1+\mu \right) ^{2}\mu} \,,\nn \\
\rho&=&-\frac { \sqrt{ \left( 1-\lambda\,\mu \right)  \left( \lambda-
\mu \right) } \left( 1+{\mu}^{2}+ \left( -4\,\lambda+2 \right) \mu
 \right)  \left( \mu-1 \right) }{ 2\left( 1+\mu \right) {\mu}^{3/2}
 \left( \lambda-1 \right) ^{2}} \,.
\eea
According to \cite{Chow:2008dp} one can read the central charge of the dual CFT of the near horizon of the extremal geometry
\be \label{ccur}
c_{\phi}=\frac{3}{2\pi}f^{\phi}\!\int dx d\phi d\psi \sqrt{\beta(x) \gamma \delta(x)}=\frac {48\pi \,{\mu}^{3/2}{k}^{3} \sqrt{1-{\xi}^{2}} \left(
1-\mu \right)  \sqrt{2\,\xi} \left( 1+\xi \right) }{ \left( 1+
\mu \right) ^{3} \left( \xi-1 \right) ^{2} \sqrt{1-2\,\xi\,
\mu+{\mu}^{2}}}\,.
\ee
The Frolov-Thorne temperature is equal to $T_{\phi}^{FT}=\frac{1}{2\pi}$, hence the microscopic Cardy entropy can be computed as
\be
S_{micro}=\frac{\pi^2}{3}c_{\phi} T_{\phi}^{FT}= \frac {8\pi^2 \,{\mu}^{3/2}{k}^{3} \sqrt{1-{\xi}^{2}} \left(
1-\mu \right)  \sqrt{2\,\xi} \left( 1+\xi \right) }{ \left( 1+
\mu \right) ^{3} \left( \xi-1 \right) ^{2} \sqrt{1-2\,\xi\,
\mu+{\mu}^{2}}}\,,
\ee
which this is equal to the macroscopic entropy when the value of the entropy in equation (\ref{tur}) evaluated at the extremal limit $\mu=\sigma$.

The central charge (\ref{ccur}) vanishes when we approach to $\xi=1$ plane (AC line) as (\ref{xiup}).
In fact this is an expected result due to the fact that on $\xi=1$ the entropy vanishes while $T_{\phi}^{FT}$ has a constant value.
By inserting the balance condition (\ref{balance}) and $\mu=\lambda/2$, which is the relation between parameters of unbalanced and balanced rings in the
extremal limit, one exactly  recovers the central charge and entropy (\ref{entem}) of the Pomeransky-Sen'kov black ring \cite{Ghodsi:2013soa,Chen:2012yd}.
%%%%%%%%%%%%%%%%%%%%%%%%%%%%%%%%%%%%%%%%%%%%%%%%%%%%%%%%%%%%%%%%%%%%%%%%%
\subsection{The double rotating dipole ring}
As in the previous case, we discuss the 2d CFT dual to the EVH cases and then analyze the extremal, but non-EVH case.

\subsubsection{The EVH dipole ring}

As discussed on the MNP plane we have a collapsing ring which is mapped onto an extremal MP black hole \eqref{MPbh} with mass and spins (\ref{d2rmp}). In the $a_1=0$ this black hole becomes EVH \cite{Ghodsi:2013soa}. It has been shown in \cite{Lu:2008jk} that there is a chiral 2d CFT associated with the near horizon geometry of this extremal solution with the following central charge
\be \label{ccmpc}
c_{CFT}=\frac{3\pi a_1^3}{2}=12\pi \frac{(1+c)^{3/2}}{\eta^{3/2}}\,.
\ee
On the other hand the near horizon metric in the EVH limit is given in (\ref{colevh}). So the Brown-Hennueax central charge is equal to
\be \label{bhcccol}
c_{BH}=\frac{3\ell}{2G_3}=12\pi \frac{(1+c)^{3/2}}{\eta^{3/2}}\,,
\ee
which is equal to (\ref{ccmpc}), supporting the EVH/CFT proposal.
\subsubsection{ The extremal solution at corner P}
In this case  it is possible to find the central charge of the dual 2d CFT as well. To this end, we note that the extremal solution (defined at $a,c\to0,\ b\to 1$) is parameterized by
$$
\alpha=\frac{c}{2a}\,,\qquad \beta=\frac{c}{1-b}.
$$
Sparing the straightforward but tedious algebra, the near horizon geometry is obtained to be
\be \label{nhedrdr}
ds^2=A(x)(-r^2dt^2+\frac{dr^2}{r^2})+B(x)dx^2+C(x)d\psi^2+D(x)[(d\phi+rdt)+\rho \,d\psi]^2 \,,
\ee
where
\bea
A(x)&=&\,{\frac {2 ({x}^{2}+1)^{1/3} \left( 1-\alpha+\alpha{x}^{2} \right) ^{2/3}{k}^{2}{\alpha}^{4/3}{\beta}^{2}}{({\alpha}^{2}-{\beta}^{2})}} \,,\nn \\
B(x)&=&\,{\frac {k^2 \alpha^{4/3} \beta^2 (\alpha x^4-\alpha+1+x^2)}{(\alpha^2-\beta^2)(x^2-1)(x^2+1)^{2/3} (1-\alpha+\alpha x^2)^{1/3}}} \,, \nn \\
C(x)&=&\,{\frac { \left( 2\alpha+1 \right) {k}^{2} \left( \alpha+\beta \right) ({x}^{2}+1)^{1/3}}{{\alpha}^{2/3} (1-\alpha+\alpha{x}^{2})^{1/3} \left( \alpha-\beta \right) }}\,,\nn \\
D(x)&=&\,{\frac {4 \left( 1-{x}^{2} \right) {k}^{2}{\alpha}^{4/3}{\beta}^{2}}{ \left( {x}^{2}+1 \right) ^{2/3}(1-\alpha+
\alpha{x}^{2})^{1/3} \left( {\alpha}^{2}-{\beta}^{2} \right) }}\,,
\nn \\
 \rho&=&\frac { \left( 2\,\alpha+1 \right) \beta+\alpha}{2\alpha \beta}\,.
\eea
The central charge can be read from the near horizon geometry (\ref{nhedrdr}) as
\be\label{ccextd2r}
c_{\phi}=\frac{3}{2\pi}\int dx d\phi d\psi \sqrt{B(x) C(x)D(x)}=\frac {24 \sqrt {2}\pi \,\alpha{k}^{3}{\beta}^{2}\sqrt {1+2\,\alpha}}{ \left( \alpha-\beta \right) \sqrt{\alpha^2-\beta^2} }, \qquad c_{\psi}=0\,.
\ee
The microscopic Cardy entropy of the dual CFT can be calculated using the fact that the Frolov-Thorne temperature is given by $T_{\phi}^{FT}=\frac{1}{2\pi}$ so the entropy will be
\be
S=\frac{\pi^2}{3}c_{\phi}T_{\phi}=\frac {4 \sqrt {2}\pi^2 \,\alpha{k}^{3}{\beta}^{2}\sqrt {1+2\,\alpha}}{ \left( \alpha-\beta \right) \sqrt{\alpha^2-\beta^2} }\,.
\ee
This is in complete agreement with the Bekenstein-Hawking entropy in (\ref{termdrd}) when it is written in terms of new parameters $\alpha$ and $\beta$
\be
S_{BH}=\frac {4 \sqrt {2}\pi^2 \,\alpha{k}^{3}{\beta}^{2}\sqrt {1+2\,\alpha}}{ \left( \alpha-\beta \right) \sqrt{\alpha^2-\beta^2} }\,.
\ee

\section{Discussion}

In this work we extended our analysis of \cite{Ghodsi:2013soa} to a larger class of  black rings. We carefully analyzed parameter space of three type of rings, the balanced and unbalanced double rotating black rings and the double rotating dipole black ring, and explored where in their parameter space they become extremal and EVH.

We found that generically the ring size parameter (denoted by $k$ or $R$, respectively for unbalanced rings and dipole rings) goes to zero in the EVH case.
Nonetheless, one can scale the other parameters to keep the mass and one spin along the ring direction finite, while the other spin (on the topologically $S^2$ part of the horizon) vanishes. Vanishing of the horizon area then comes from the vanishing of a one-cycle along the $S^2$ part. This direction parameterized by $\phi$ joins time and radial direction to form a (pinching) AdS$_3$ factor in the near horizon limit. As we explicitly showed the near horizon geometry of all EVH rings, balanced, unbalanced or dipole-charged, becomes precisely the same geometry, the one which is also obtained as the near horizon limit of EVH MP black hole with the same mass as the EVH rings. The AdS$_3$ radius for all these cases becomes $\ell^2=\frac{8M}{3\pi}$, where $M$ is the mass of the EVH black ring/hole.\footnote{In the unbalanced case there remains a trace of the unbalance factor in the range of coordinate parameterizing the pinching direction $\phi$.} This result is of course understandable because all these near horizon geometries are 5d vacuum Einstein solutions. (Note that in the dipole ring case, the dipole charge goes to zero in the near horizon limit of EVH ring.) And there is a uniqueness theorem for such solutions with $SO(2,2)$ isometry \cite{SO22-uniqueness}.

For the near-EVH rings, in the near horizon limit the AdS$_3$ factor turns into a BTZ black hole. The mass and angular momentum of the BTZ exactly captures the deviations of the mass and angular momenta of the near-EVH black ring from the EVH point. In other words, these near-EVH excitations survive the near horizon limit. One may view this fact as the zeroth order evidence in support of the EVH/CFT proposal which states that all excitations or perturbations around an EVH black hole/ring is governed by a 2d CFT dual to the AdS$_3$ throat appearing in the near horizon geometry, or dual to the 3d gravity obtained from reducing the 5d theory over the $\theta\hat\psi$ part of the near horizon geometry.

As is discussed in the literature and we reviewed here, near the collapsing regions of the black ring parameter space one may always find a new coordinate system where the ring solution is mapped onto a MP black hole. Moreover, this collapsing region generically intersects the extremal surface in the parameter space
(for example see Fig.(\ref{fig1}.a) and Fig.(\ref{fig2})). The ring-hole map at these intersections, as expected, then relates an extremal ring to an extremal hole. On the other hand, we showed that the EVH rings appear around this collapsing regions. In general near-EVH ring we get a BTZ metric in the near horizon limit, while around these intersection points/lines we get an extremal BTZ in the near horizon.
Given this picture, EVH rings provide the window that ring-hole transition may occur. This transition may be traced from the 2d CFT which is  proposed to capture the low energy dynamics of the EVH rings. We hope to study this point further in future works.

%%%%%%%%%%%%%%%%%%%%%%%%%%%%%%%%%%%%%%%%%%%%%%%%%%%%%%%%%%%%%%%%%%%%%%%%%%%%%%%

\section*{Acknowledgment}
We would like to thank Roberto Emparan and Hossein Yavartanoo for comments on the draft. A.G. and H.G. would like to thank the IPM for hospitality while this project was completed. H.G. would also like to thank Davood Mahdavian Yekta and Ghadir Jafari for useful discussions.
The work of A.G. and H.G. is supported by Ferdowsi University of Mashhad under the grant
2/31025 (1393/04/10).

 %%%%%%%%%%%%%%%%%%%%%%%%%%%%%%%%%%%%%%%

%\endgroup


\begin{thebibliography}{99}

\bibitem{MP}
  R.~C.~Myers and M.~J.~Perry,
 ``Black Holes in Higher Dimensional Space-Times,''
  Annals Phys.\  {\bf 172}, 304 (1986).
  %%CITATION = APNYA,172,304;%%

  \bibitem{Emparan:2001wn}
  R.~Emparan and H.~S.~Reall,
  ``A Rotating black ring solution in five-dimensions,''
  Phys.\ Rev.\ Lett.\  {\bf 88}, 101101 (2002)
  [hep-th/0110260].

   %\cite{Pomeransky:2006bd}
\bibitem{Pomeransky:2006bd}
  A.~A.~Pomeransky and R.~A.~Sen'kov,
  ``Black ring with two angular momenta,''
  hep-th/0612005.

  \bibitem{Emparan:2008eg}
  R.~Emparan and H.~S.~Reall,
  ``Black Holes in Higher Dimensions,''
  Living Rev.\ Rel.\  {\bf 11}, 6 (2008)
  [arXiv:0801.3471 [hep-th]].
  %%CITATION = ARXIV:0801.3471;%%


\bibitem{Elvang-Emparan-2003}
  H.~Elvang, R.~Emparan and ,
``Black rings, supertubes, and a stringy resolution of black hole nonuniqueness,''
  JHEP {\bf 0311} (2003) 035
  [hep-th/0310008].
  %%CITATION = HEP-TH/0310008;%%


\bibitem{Emparan-Obers-2007}
  R.~Emparan, T.~Harmark, V.~Niarchos, N.~A.~Obers and M.~J.~Rodriguez,
  ``The Phase Structure of Higher-Dimensional Black Rings and Black Holes,''
  JHEP {\bf 0710} (2007) 110
  [arXiv:0708.2181 [hep-th]].
  %%CITATION = ARXIV:0708.2181;%%

\bibitem{ur}
Y.~Morisawa, S.~Tomizawa and Y.~Yasui,
Boundary Value Problem for Black Rings,
  Phys.\ Rev.\ D {\bf 77}, 064019 (2008)
  [arXiv:0710.4600 [hep-th]].

\bibitem{A-R-T-papers}
D.~Astefanesei, M.~J.~Rodriguez and S.~Theisen,
``Quasilocal equilibrium condition for black ring,''
  JHEP {\bf 0912} (2009) 040
  [arXiv:0909.0008 [hep-th]];
  %%CITATION = ARXIV:0909.0008;%%
%D.~Astefanesei, M.~J.~Rodriguez and S.~Theisen,
``Thermodynamic instability of doubly spinning black objects,''
  JHEP {\bf 1008} (2010) 046
  [arXiv:1003.2421 [hep-th]].
  %%CITATION = ARXIV:1003.2421;%%

\bibitem{Chen:2011jb}
  Y.~Chen, K.~Hong and E.~Teo,
  ``Unbalanced Pomeransky-Sen'kov black ring,''
  Phys.\ Rev.\ D {\bf 84}, 084030 (2011)
  [arXiv:1108.1849 [hep-th]].



\bibitem{dipole}
  R.~Emparan,
 ``Rotating circular strings, and infinite nonuniqueness of black rings,''
  JHEP {\bf 0403}, 064 (2004)
  [hep-th/0402149].
  %%CITATION = HEP-TH/0402149;%%

\bibitem{Yazadjiev}
  S.~S.~Yazadjiev,
``Completely integrable sector in 5-D Einstein-Maxwell gravity and derivation of the dipole black ring solutions,''
  Phys.\ Rev.\ D {\bf 73}, 104007 (2006)
  [hep-th/0602116]; ``Solution generating in 5D Einstein-Maxwell-dilaton gravity and derivation of dipole black ring solutions,''
  JHEP {\bf 0607}, 036 (2006)
  [hep-th/0604140].
  %%CITATION = HEP-TH/0604140;%%
  %%CITATION = HEP-TH/0602116;%%


\bibitem{Chen:2012kd}
  Y.~Chen, K.~Hong and E.~Teo,
  ``A Doubly rotating black ring with dipole charge,''
  JHEP {\bf 1206}, 148 (2012)
  [arXiv:1204.5785 [hep-th]].
  %%CITATION = ARXIV:1204.5785;%%

\bibitem{Sudarsky-Wald}
D. Sudarsky and R. M. Wald, ``Extrema of mass, stationarity, and static-ity, and solutions to the Einstein Yang-Mills equations'' Phys. Rev. {\bf D46}, 1453 (1992); R. M. Wald, ''The First law of black hole mechanics,'' arXiv:gr-qc/9305022.

\bibitem{Horowitz-Copsey}
  K.~Copsey and G.~T.~Horowitz,
``The Role of dipole charges in black hole thermodynamics,''  Phys.\ Rev.\ D {\bf 73}, 024015 (2006)  [hep-th/0505278].  %%CITATION = HEP-TH/0505278;%%


 \bibitem{Ghodsi:2013soa}
  A.~Ghodsi, H.~Golchin and M.~M.~Sheikh-Jabbari,
  ``Dual 2d CFT Identification of Extremal Black Rings from Holes,'' JHEP {\bf 1310}, 194 (2013)
  arXiv:1308.1478 [hep-th].

\bibitem{SheikhJabbaria:2011gc}
  M.~M.~Sheikh-Jabbari and H.~Yavartanoo,
  ``EVH Black Holes, AdS3 Throats and EVH/CFT Proposal,''
  JHEP {\bf 1110}, 013 (2011)
  [arXiv:1107.5705 [hep-th]].
  %%CITATION = ARXIV:1107.5705;%%

\bibitem{EVH-examples}
R.~Fareghbal, C.~N.~Gowdigere, A.~E.~Mosaffa and M.~M.~Sheikh-Jabbari,
``Nearing Extremal Intersecting Giants and New Decoupled Sectors in N = 4 SYM,''
  JHEP {\bf 0808}, 070 (2008)
  [arXiv:0801.4457 [hep-th]]; ``Nearing 11d Extremal Intersecting Giants and New Decoupled Sectors in D = 3,6 SCFT's,''
  Phys.\ Rev.\ D {\bf 81}, 046005 (2010)
  [arXiv:0805.0203 [hep-th]].

T.~Azeyanagi, N.~Ogawa and S.~Terashima,
  ``Emergent AdS$_3$ in the Zero Entropy Extremal Black Holes,''
  JHEP {\bf 1103} (2011) 004
  [arXiv:1010.4291 [hep-th]].
  %%CITATION = ARXIV:1010.4291;%%


  H.~Yavartanoo,
``On heterotic black holes and EVH/CFT correspondence,''
  Eur.\ Phys.\ J.\ C {\bf 72}, 2256 (2012); ``EVH Black Hole Solutions With Higher Derivative Corrections,''
  Eur.\ Phys.\ J.\ C {\bf 72}, 1911 (2012)
  [arXiv:1301.4174 [hep-th]]; ``On EVH black hole solution in heterotic string theory,''
  Nucl.\ Phys.\ B {\bf 863}, 410 (2012)
  [arXiv:1212.3742 [hep-th]]; ``Five-dimensional heterotic black holes and its dual IR-CFT,''
  Eur.\ Phys.\ J.\ C {\bf 72} (2012) 2197 [arXiv:1301.3706 [physics.gen-ph]].

\bibitem{deBoer:2011zt}
 J.~de Boer, M.~Johnstone, M.~M.~Sheikh-Jabbari and J.~Simon,
 ``Emergent IR Dual 2d CFTs in Charged AdS5 Black Holes,''
  Phys.\ Rev.\ D {\bf 85}, 084039 (2012)
  [arXiv:1112.4664 [hep-th]].

M.~Johnstone, M.~M.~Sheikh-Jabbari, J.~Simon and H.~Yavartanoo,
``Near-Extremal Vanishing Horizon AdS5 Black Holes and Their CFT Duals,''
  JHEP {\bf 1304}, 045 (2013)
  [arXiv:1301.3387 [hep-th]].

\bibitem{Johnstone:2013ioa}
  M.~Johnstone, M.~M.~Sheikh-Jabbari, J.~Simon and H.~Yavartanoo,
  ``Extremal Black Holes and First Law of Thermodynamics,''
Phys.\ Rev.\ D {\bf 88}, no. 10, 101503 (2013)  arXiv:1305.3157 [hep-th].

\bibitem{deBoer:2010ac}
  J.~de Boer, M.~M.~Sheikh-Jabbari and J.~Simon,
 ``Near Horizon Limits of Massless BTZ and Their CFT Duals,''
  Class.\ Quant.\ Grav.\  {\bf 28}, 175012 (2011)
  [arXiv:1011.1897 [hep-th]].

\bibitem{Gubser}
  S.~S.~Gubser,
``Curvature singularities: The Good, the bad, and the naked,''
  Adv.\ Theor.\ Math.\ Phys.\  {\bf 4}, 679 (2000)
  [hep-th/0002160].
  %%CITATION = HEP-TH/0002160;%%

\bibitem{Elvang:2007hs}
  H.~Elvang and M.~J.~Rodriguez,
  ``Bicycling Black Rings,''
  JHEP {\bf 0804} (2008) 045
 [arXiv:0712.2425 [hep-th]].


\bibitem{SO22-uniqueness}
S. Sadeghian, M.M. Sheikh-Jabbari, H. Yavartanoo, ``On  Classification of Geometries with SO(2,2) Symmetry,'' \emph{To appear}.



\bibitem{Figueras:2005zp}
  P.~Figueras,
  ``A Black ring with a rotating 2-sphere,''
  JHEP {\bf 0507}, 039 (2005)
  [hep-th/0505244].


\bibitem{Guica:2008mu}
  M.~Guica, T.~Hartman, W.~Song and A.~Strominger,
  ``The Kerr/CFT Correspondence,''
  Phys.\ Rev.\ D {\bf 80}, 124008 (2009)
  [arXiv:0809.4266 [hep-th]].
  %%CITATION = ARXIV:0809.4266;%%


  %\cite{Chen:2012yd}
\bibitem{Chen:2012yd}
 B.~Chen and J.~-j.~Zhang,
  ``Holographic Descriptions of Black Rings,''
  JHEP {\bf 1211}, 022 (2012)
  [arXiv:1208.4413 [hep-th]].
  %%CITATION = ARXIV:1208.4413;%%

\bibitem{BH}J. D. Brown and M. Henneaux, {\it Central Charges in the Canonical Realization of
 Asymptotic Symmetries: An Example from Three-Dimensional Gravity},
 Commun. Math. Phys. 104 (1986) 207�226;

\bibitem{Chow:2008dp}
  D.~D.~K.~Chow, M.~Cvetic, H.~Lu and C.~N.~Pope,
  ``Extremal Black Hole/CFT Correspondence in Gauged Supergravities,''
  Phys.\ Rev.\ D {\bf 79}, 084018 (2009)
  [arXiv:0812.2918 [hep-th]].
\bibitem{Lu:2008jk}
  H.~Lu, J.~Mei and C.~N.~Pope,
  ``Kerr-AdS/CFT Correspondence in Diverse Dimensions,''
  JHEP {\bf 0904}, 054 (2009)
  [arXiv:0811.2225 [hep-th]].


\end{thebibliography}
 \end{document}